\documentclass[11pt]{article}
\usepackage{axodraw}
\usepackage{mathtext}         %

\usepackage{amsmath,amssymb} 
\usepackage{epsfig}

\newcommand{\be}{\begin{equation}}
\newcommand{\ee}{\end{equation}}
\newcommand{\bea}{\begin{eqnarray}}
\newcommand{\eea}{\end{eqnarray}}
\newcommand{\ba}{\begin{eqnarray*}}
\newcommand{\ea}{\end{eqnarray*}}

\newcommand{\Fe}[2]{\Fs{#1}{#2}{z}}
\newcommand{\Fw}[2]{\Fs{#1}{#2}{w}}

\newcommand{\FwZ}[2]{\Fs{#1}{#2}{ w z }}

\newcommand{\FBX}[2]{\Fs{#1}{#2}{ \frac{y}{x+y-xy} }}

\newcommand{\FIP}[2]{\Fs{#1}{#2}{\frac{4(s_{35}+s_{14}-s_{13}) }{s_{35}s_{14}h_5} }}
\newcommand{\FDB}[2]{\Fs{#1}{#2}{\frac{4(s_{14}+s_{25}-s_{24}) }{s_{25}s_{14}h_5} }}

\newcommand{\FhP}[2]{\Fs{#1}{#2}{\frac{4}{s_{25}\overline{h}_5} }}
\newcommand{\FhC}[2]{\Fs{#1}{#2}{\frac{4}{s_{24}\overline{h}_5} }}
\newcommand{\FBR}[2]{\Fs{#1}{#2}{\frac{-4s_{13}}{s_{14} s_{35}\overline{h}_5} }}

\newcommand{\Fiw}[2]{\Fs{#1}{#2}{\frac{1}{w}}}

\newcommand{\Fh}[2]{\,{}_#1F_#2}

\newcommand{\Fs}[3]{\!\!\left[\begin{array}{c}#1\,;\\#2\,;\end{array}#3\right]}

\newcommand{\FSa}[2]{\Fs{#1}{#2}{\frac{s_{35}+s_{24}-s_{25}}{s_{35}} }}

\newcommand{\FCP}[2]{\Fs{#1}{#2}{\frac{s_{24}}{s_{25}} }}

\newcommand{\FSd}[2]{\Fs{#1}{#2}{\frac{s_{24}(s_{14}+s_{25}-s_{24})}{s_{14}s_{25}} }}

\newcommand{\FSA}[2]{\Fs{#1}{#2}{\frac{s(c+b-a)}{b c} }}
\newcommand{\FSC}[2]{\Fs{#1}{#2}{\frac{s(s_{13}+s_{24}-s_{14})}{s_{13}s_{24}} }}
\newcommand{\FIR}[2]{\Fs{#1}{#2}{\frac{s_{13}s_{24}}{s(s_{13}+s_{24}-s_{14})} }}

\newcommand{\FPD}[2]{\Fs{#1}{#2}{\frac{s_{24}}{s_{35}+s_{24}-s_{25} } }}

\newcommand{\FPH}[2]{\Fs{#1}{#2}{\frac{s_{14}}{s_{14}+s_{25}-s_{24} } }}

\newcommand{\FPF}[2]{\Fs{#1}{#2}{\frac{s_{24}s_{35}}{s_{25}(s_{35}+s_{24}-s_{25}) } }}
\newcommand{\FPC}[2]{\Fs{#1}{#2}{\frac{s_{25}}{s_{14}+s_{25}-s_{24} } }}
\newcommand{\FPE}[2]{\Fs{#1}{#2}{\frac{s_{25}}{s_{35}} }}
\newcommand{\FPG}[2]{\Fs{#1}{#2}{\frac{s_{24}}{s_{14}} }}
\newcommand{\Fx}[2]{\Fs{#1}{#2}{x}}
\newcommand{\Fy}[2]{\Fs{#1}{#2}{y}}
\newcommand{\FYM}[2]{\Fs{#1}{#2}{ \frac{y}{x(y-1)}   }}
\newcommand{\FYC}[2]{\Fs{#1}{#2}{ \frac{y}{y-1}   }}
\newcommand{\FXM}[2]{\Fs{#1}{#2}{ \frac{y(x-1)}{x}   }}

\newcommand{\FSD}[2]{\Fs{#1}{#2}{\frac{-4s_{24}s_{25}}{(s_{24}-s_{25})^2 }}}

\newcommand{\Fuy}[2]{\Fs{#1}{#2}{ u y}}

\newcommand{\FZA}[2]{\Fs{#1}{#2}{\frac{s_{35}s_{14}}{(s_{13}-s_{35})(s_{13}-s_{14})}
}}
\newcommand{\FZB}[2]{\Fs{#1}{#2}{\frac{s_{25}}{s_{35}-s_{13}}}}
\newcommand{\FZC}[2]{\Fs{#1}{#2}{\frac{s_{35}}{s_{13}-s_{14}}}}
\newcommand{\FZD}[2]{\Fs{#1}{#2}{- \frac{s_{25}}{s_{14}}}}
\newcommand{\FZE}[2]{\Fs{#1}{#2}{- \frac{s_{14}}{s_{25}}}}
\newcommand{\FZF}[2]{\Fs{#1}{#2}{ \frac{s_{25}+s_{13}-s_{35} }{s_{13}}}}
\newcommand{\FZG}[2]{\Fs{#1}{#2}{\frac{s_{14}}{s_{13}-s_{35}}}}
\newcommand{\FZH}[2]{\Fs{#1}{#2}{\frac{s_{13}s_{25}}{(s_{25}-s_{35})(s_{13}-s_{35})}
}}

\begin{document}

\begin{titlepage}
\thispagestyle{empty}
\onecolumn

\begin{flushright}
\today  \\
DESY~09--218\\
\end{flushright}

\vspace*{0.2cm}

\begin{center}

{\bf \Large Analytic result
for the one-loop scalar pentagon \\
\vspace{3mm}
  integral with massless propagators
}

\vspace{12mm}

\vspace{12mm}

                {Bernd A. Kniehl, Oleg V. Tarasov\footnote{On leave
of absence from Joint Institute for Nuclear Research,
141980 Dubna (Moscow Region) Russia.}        \\
{\normalsize\it II. Institut f\"ur Theoretische Physik, 
Universit\"at Hamburg,}\\
{\normalsize\it Luruper Chaussee 149, 22761 Hamburg, Germany}}

\end{center}

\vspace*{1.0cm}

\begin{abstract}
The method of dimensional recurrences proposed by one of the authors 
\cite{Tarasov:1996br, Tarasov:2000sf} is applied to the evaluation of the
pentagon-type
scalar integral with on-shell external legs and massless internal lines.
For the first time, an analytic result valid for arbitrary  space-time
dimension $d$ and five arbitrary kinematic variables  is presented. 
An explicit expression in terms of the Appell hypergeometric function $F_3$ 
and the Gauss hypergeometric function ${_2F_1}$, both admitting one-fold
integral representations, is given.
In the case when one kinematic variable vanishes, the integral reduces
to a combination of Gauss hypergeometric functions ${_2F_1}$.
For the case when one scalar invariant is large compared to the others,
the asymptotic values of the integral in terms of Gauss hypergeometric
functions ${_2F_1}$  are presented in $d=2-2\varepsilon$, $4-2\varepsilon$,  
and $6-2\varepsilon$ dimensions. For multi-Regge kinematics, the 
asymptotic value of the integral in $d=4-2\varepsilon$ dimensions is given 
in terms of the Appell function $F_3$ and the Gauss hypergeometric 
function ${_2F_1}$.

\medskip

\medskip

\medskip


\end{abstract}
\end{titlepage}

\newpage

\section{Introduction}

Theoretical predictions for ongoing and future experiments at the CERN Large
Hadron Collider (LHC) and an International $e^+e^-$ Linear Collider (ILC) must
include high-precision radiative corrections.
The complexity of the evaluation of such radiative corrections is related, in
particular, to the difficulties in calculating integrals corresponding to
Feynman diagrams with many external legs depending on many kinematic
variables.
Purely numerical evaluation of such integrals cannot provide sufficiently high
precision within reasonable computer time.
The evaluation of one-loop integrals corresponding to diagrams with two, three,
and four external legs was studied in numerous publications.

As for integrals associated with diagrams with five and more external legs, the
situation is quite different.
Not so many results for such integrals are available in the literature.  
Various authors \cite{brown} discussed the reduction of pentagon integrals to
box integrals in space-time dimension $d=4$.
They were able to express these integrals as linear combinations of five
different loop integrals with four external legs.
Infrared divergences if any were supposed to be regulated by introducing a
small fictitious mass.
The representation of the dimensionally regularized pentagon integral in terms
of box-type integrals was considered in Refs.~\cite{Bern:1992em,Bern:1993kr}.
However, in the calculation of multi-loop radiative corrections, higher orders
in $\varepsilon=(4-d)/2$ are needed, and, therefore, one should extend such an
expansion beyond the ``box approximation.''
The first step in this direction was recently taken in
Ref.~\cite{DelDuca:2009ac}, where an analytic result for the one-loop massless
pentagon integral with on-shell external legs as well as several terms of its 
$\varepsilon$ expansion were presented in the limit of multi-Regge kinematics.
However, no analytic results for arbitrary kinematics are available until now. 
Practically nothing is known about the analytic structure of on-shell
pentagon integrals with unconstrained kinematics in arbitrary space-time
dimension $d$.
For very simplified kinematics, a pentagon-type integral for arbitrary value of
$d$ was given in Ref.~\cite{Fadin:1996yv} in terms of Euler gamma functions.
A representation of the pentagon integral in terms of a four-fold Mellin-Barnes
integral may be found in Ref.~\cite{Davydychev:1990jt}.

Important applications that require the evaluation of Feynman integrals with
massless propagators include the study of jet production in QCD
\cite{Giele:1992vf}, which allows for a high-precision extraction of the
strong-coupling constant $\alpha_s$, the investigation of the iterative 
structure of ${\cal N}=4$ supersymmetric Yang-Mills (SYM) amplitudes
\cite{Bern:2005iz}, and tests of the scattering-amplitude/Wilson-loop duality
\cite{Drummond:2007aua}.

In this paper, we perform a first analytic study of the on-shell pentagon
integral with massless internal lines and arbitrary kinematic invariants.
We use the method of dimensional recurrences proposed in
Refs.~\cite{Tarasov:1996br,Tarasov:2000sf}, which was already applied to the
calculation of one- and two-loop integrals in
Refs.~\cite{Tarasov:2000sf,Fleischer:2003rm,Tarasov:2006nk}
and, quite recently, also to the calculation of three- and four-loop integrals
in Ref.~\cite{Lee:2009dh}.

This paper is organized as follows.
In Section~2, we introduce our notations, explain the recurrence relation for
the pentagon-type integral with respect to the space-time dimension $d$, and
we present a detailed derivation of its solution in Section~3. 
In Section~4, we consider a particular case of the on-shell pentagon integral
with one vanishing kinematic variable and present an analytic expression in
terms of the Gauss hypergeometric function ${_2F_1}$.
In Section~5, we specify the asymptotic values of the pentagon integral in
$d=2-2\varepsilon$, $d=4-2\varepsilon$, and $d=6-2\varepsilon$ space-time
dimensions when one of the scalar invariants is much larger than the others.
In Section~6, we present an analytic expression for the pentagon integral in
the limit of multi-Regge kinematics in terms of the Appell function $F_3$ and
the Gauss hypergeomteric function ${_2F_1}$. 
In the Conclusions, we summarize the accomplishments of the present paper and
offer some perspectives for the application of the method of dimensional
recurrences to six-point integrals.
In Appendix~A, we collect useful formulae for hypergeometric functions used
in this paper.
Appendix~B contains intermediate results from Section~3.


\section{Definitions and dimensional recurrences}


We consider the following integral with five massless propagators:
\begin{equation}
I_5^{(d)}(s_{12},s_{23},s_{34},s_{45},
          s_{15};~s_{13},s_{14},
          s_{24},s_{25},
          s_{35})
=\int \frac{d^dq}{i \pi^{d/2}} \prod_{j=1}^5 \frac{1}{D_j},
\label{i5_definition}
\end{equation}
where 
\begin{equation}
D_j=(q-p_j)^2+i \epsilon.
\label{propagators}
\end{equation}
The pentagon diagram associated with the integral $I_5^{(d)}(\{s_{kr}\})$ is
depicted in Fig.~1. 
The labeling of the momenta in Fig.~1. corresponds to
Eq.~(\ref{i5_definition}).
The Lorentz invariants are defined as:
\begin{equation}
  \qquad  s_{ij}=p_{ij}^2, 
  \qquad  p_{ij}=p_i-p_j.
\end{equation}
In the present paper, we take the squares of the external momenta to be
vanishing,
\begin{equation}
s_{12}=s_{23}=s_{34}=s_{45}=s_{15}=0,
\label{onshell}
\end{equation}
and work in the Euclidean region, $s_{ij}<0$.
In what follows, we only keep non-vanishing variables as arguments of
$I_5^{(d)}$ and use the notation
\begin{equation}
 I_5^{(d)}(s_{13},s_{14},s_{24},s_{25},s_{35})\equiv
I_5^{(d)}(0,0,0,0,
          0;~s_{13},s_{14},
          s_{24},s_{25},
          s_{35}),
\end{equation}
for the on-shell case of Eq.~(\ref{onshell}). 
\vspace{1cm}
\begin{figure}[h]
\begin{center}
\includegraphics[scale=1.1]{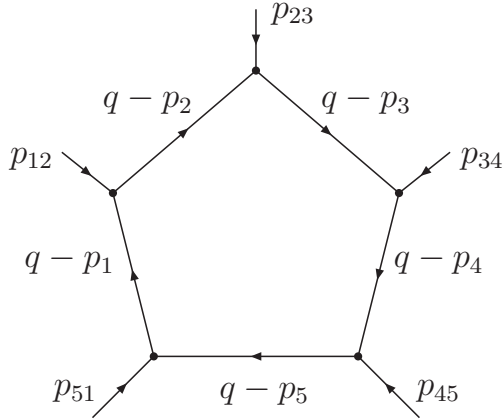}
\end{center}
\caption{Feynman diagram corresponding the integral $I_5^{(d)}$.}
\end{figure}
\vspace {0.5cm}
Due to the symmetry with respect to permutations of all propagators, the
integral considered as a function of the kinematic variables $s_{ij}$ must
fulfill the following  relations:
\begin{eqnarray}
&&           I_5^{(d)}( s_{13}, s_{14}, s_{24}, s_{25}, s_{35})
            = I_5^{(d)}( s_{13}, s_{35}, s_{25}, s_{24}, s_{14}) \nonumber \\
&&          = I_5^{(d)}( s_{14}, s_{13}, s_{35}, s_{25}, s_{24})
            = I_5^{(d)}( s_{14}, s_{24}, s_{25}, s_{35}, s_{13}) \nonumber \\
&&          = I_5^{(d)}( s_{24}, s_{25}, s_{35}, s_{13}, s_{14}) 
            = I_5^{(d)}( s_{24}, s_{14}, s_{13}, s_{35}, s_{25}) \nonumber \\
&&          = I_5^{(d)}( s_{25}, s_{24}, s_{14}, s_{13}, s_{35}) 
            = I_5^{(d)}( s_{25}, s_{35}, s_{13}, s_{14}, s_{24}) \nonumber \\
&&          = I_5^{(d)}( s_{35}, s_{13}, s_{14}, s_{24}, s_{25}) 
            = I_5^{(d)}( s_{35}, s_{25}, s_{24}, s_{14}, s_{13}).
\label{simmetry}
\end{eqnarray}

As was shown in Refs.~\cite{Tarasov:1996br,Bern:1993kr}, the integrals
$I_5^{(d+2)}$ and $I_5^{(d)}$ fulfill the following relation:
\begin{eqnarray}
&&
(d-4) s_{13}s_{24} s_{35} s_{14} s_{25}
 h_5 I_5^{(d+2)}(s_{13},s_{14},s_{24},s_{25},s_{35})
\nonumber \\
&& \nonumber \\
&&~~~~~~~~~~~~~~~~~~~~~~~~~
 =
 2 s_{13}s_{24} s_{35} s_{14} s_{25}
 I_5^{(d)}(s_{13},s_{14},s_{24},s_{25},s_{35})
+P^{(d)},
\label{dim_rec_i5}
\end{eqnarray}
where
\begin{equation}
P^{(d)}=\sum_{k=1}^5 \varkappa_k J_k,
\end{equation} 
\begin{eqnarray}
&&\varkappa_1=
- s_{24}s_{35}(s_{24}s_{35}-s_{35}s_{14}-s_{13}s_{24}+s_{13}s_{25}
+s_{14}s_{25}),
\nonumber \\
&&\varkappa_2=
-s_{35} s_{14}(s_{35} s_{14}+s_{13}s_{24}-s_{24}s_{35}-s_{14}s_{25}+s_{13}s_{25}),
\nonumber \\
&&\varkappa_3=
-s_{14}s_{25}(s_{14}s_{25}-s_{35}s_{14}-s_{13} s_{25}+s_{24} s_{35}+s_{13} s_{24}) ,
\nonumber \\
&&\varkappa_4=
s_{13} s_{25} (s_{13} s_{24}-s_{13} s_{25}-s_{24} s_{35}-s_{35} s_{14}+s_{14} s_{25}),
\nonumber \\
&&\varkappa_5=
s_{13} s_{24} (s_{13} s_{25}-s_{13} s_{24}-s_{35} s_{14}
+s_{24} s_{35}-s_{14} s_{25}), 
\label{P_ot_d}
\end{eqnarray}
\begin{eqnarray}
&&J_1=I_4^{(d)}(0,0,0,s_{25};~s_{24},s_{35}),~~~~J_2=I_4^{(d)}(0,0,0,s_{13};~s_{14},s_{35}),
\nonumber \\
&&J_3=I_4^{(d)}(0,0,0,s_{24};~s_{25},s_{14}),~~~~J_4=I_4^{(d)}(0,0,0,s_{35};~s_{13},s_{25}),
\nonumber \\
&&J_5=I_4^{(d)}(0,0,0,s_{14};~s_{24},s_{13}),
\label{I4integrals}
\end{eqnarray}
\begin{eqnarray}
h_5 &=&\frac{1}{s_{13}s_{24}s_{35}s_{14}s_{25}}\left[
s_{24}^2 s_{35}^2-2 s_{24} s_{13}^2 s_{25}+s_{13}^2 s_{24}^2
+s_{13}^2 s_{25}^2+s_{14}^2 s_{35}^2+s_{14}^2 s_{25}^2
\right.
\nonumber \\
&&
+2 s_{13} s_{24} s_{35} s_{14}
-2 s_{24} s_{14} s_{35}^2
-2 s_{13} s_{35} s_{24}^2+2 s_{24} s_{13} s_{35} s_{25}
+2 s_{24} s_{35} s_{14} s_{25}
\nonumber \\
&&\left.
-2 s_{35} s_{14}^2 s_{25}+2 s_{13} s_{35} s_{14} s_{25}
-2 s_{13} s_{14} s_{25}^2+2 s_{24} s_{13} s_{14} s_{25}
\right].
\end{eqnarray}
Here, $I_4^{(d)}$ are integrals corresponding to Feynman diagrams with four
external legs defined as
\begin{equation}
I_4^{(d)}(s_{nj},s_{jk},s_{kl},s_{nl};~s_{jl},s_{nk})
=
\int \frac{d^dq}{i\pi^{d/2}}
~\frac{1}{ D_n D_j D_k D_l},
\end{equation}
where $D_j$ are defined in Eq.~(\ref{propagators}).
An analytic expression for this integral with arbitrary kinematics in arbitrary
space-time dimension $d$ was recently obtained in Ref.~\cite{Kniehl:2009b}.


\section{Solution of the dimensional recurrence relation}


Redefining the integral $I_5^{(d)}$ as
\begin{equation}
I_5^{(d)} (s_{13},s_{14},s_{24},s_{25},s_{35})=
\frac{{h_5}^{-\frac{d}{2} }}
{\Gamma\left(\frac{d-4}{2} \right)} 
\overline{I}^{(d)}_5(s_{13},s_{14},s_{24},s_{25},s_{35}),
\label{i5_redefinition}
\end{equation}
we obtain the relation
\begin{equation}
\overline{I}^{(d+2)}_5(s_{13},s_{14},s_{24},s_{25},s_{35}) 
=
\overline{I}^{(d)}_5(s_{13},s_{14},s_{24},s_{25},s_{35}) 
+\frac{\Gamma\left(\frac{d-4}{2}\right)}
 { 2 s_{13}s_{24} s_{35} s_{14} s_{25}} h_5^{\frac{d}{2}}
P^{(d)},
\label{equ_4_I5bar}
\end{equation}
which has the following solution:
\begin{equation}
\overline{I}^{(d)}_5(s_{13},s_{14},s_{24},s_{25},s_{35})=
\Pi_{m}(d)+\sum_{r=0}^{\infty}
\frac{\Gamma\left(\frac{d-2r-6}{2}\right)} 
{2  s_{13}s_{24} s_{35} s_{14} s_{25}}
h_5^{\frac{d}{2}-r-1} P^{(d-2r-2)},
\label{solution_i5m}
\end{equation}
where $\Pi_m(d)$ is an arbitrary periodic function depending on the scalar
invariants $s_{ij}$ and satisfying the condition
\begin{equation}
\Pi_m(d+2)=\Pi_m(d).
\end{equation}
Another solution of Eq.~(\ref{equ_4_I5bar}) reads:
\begin{equation}
\overline{I}^{(d)}_5(s_{13},s_{14},s_{24},s_{25},s_{35})=
\Pi_{p}(d)-\sum_{r=0}^{\infty}
\frac{\Gamma\left(\frac{d+2r-4}{2}\right)}
 {2  s_{13}s_{24} s_{35} s_{14} s_{25}}
h_5^{\frac{d}{2}+r}
P^{(d+2r)},
\label{solution_i5p}
\end{equation}
where
\begin{equation}
\Pi_p(d+2)=\Pi_p(d).
\label{periodic_p}
\end{equation}
The correctness of both solutions, Eqs.~(\ref{solution_i5m}) and 
(\ref{solution_i5p}), may be easily verified by direct  substitution into 
Eq.~(\ref{equ_4_I5bar}).
Thus, for example, using Eqs.~(\ref{solution_i5p}) and (\ref{periodic_p}), we
have
\begin{eqnarray}
&&\overline{I}^{(d+2)}_5(s_{13},s_{14},s_{24},s_{25},s_{35})
= \Pi_{p}(d+2)-\sum_{r=0}^{\infty}
\frac{\Gamma\left(\frac{d+2+2r-4}{2}\right)}
 {2  s_{13}s_{24} s_{35} s_{14} s_{25}}
h_5^{\frac{d+2}{2}+r}
P^{(d+2+2r)}
\nonumber \\
&&~~~~~~~~~=\Pi_{p}(d)
-\sum_{r=1}^{\infty}
\frac{\Gamma\left(\frac{d+2r-4}{2}\right)}
 {2  s_{13}s_{24} s_{35} s_{14} s_{25}} h_5^{\frac{d}{2}+r}
 P^{(d+2r)}
\nonumber \\
&&~~~~~~~~~=\overline{I}^{(d)}_5(s_{13},s_{14},s_{24},s_{25},s_{35})
+ \frac{\Gamma\left(\frac{d-4}{2}\right)}
 {2  s_{13}s_{24} s_{35} s_{14} s_{25}} h_5^{\frac{d}{2}} P^{(d)},
\end{eqnarray}
in agreement with Eq.~(\ref{equ_4_I5bar}).
Furthermore, the solution in the form of Eq.~(\ref{solution_i5p}) may be
easily obtained from Eq.~(\ref{solution_i5m}) by adding to and subtracting from
the expression on the right-hand side of Eq.~(\ref{solution_i5m}) the sum
\begin{equation}
\sum_{r=-1}^{-\infty}
\frac{\Gamma\left(\frac{d-2r-6}{2}\right)}
 {2  s_{13}s_{24} s_{35} s_{14} s_{25}}
h_5^{\frac{d}{2}-r-1} P^{(d-2r-2)}.
\end{equation}
In fact, the combination
\begin{eqnarray}
&&
\sum_{r=0}^{\infty}
\frac{\Gamma\left(\frac{d-2r-6}{2}\right)}
 {2  s_{13}s_{24} s_{35} s_{14} s_{25}}
h_5^{\frac{d}{2}-r-1} P^{(d-2r-2)}
+
\sum^{-\infty}_{r=-1}
\frac{\Gamma\left(\frac{d-2r-6}{2}\right)}
 {2  s_{13}s_{24} s_{35} s_{14} s_{25}}
h_5^{\frac{d}{2}-r-1} P^{(d-2r-2)} 
\nonumber \\
&&~~~~~~~~~~~~~~~~~~~~~~~~~~
= \sum_{r=-\infty}^{\infty}
\frac{\Gamma\left(\frac{d-2r-6}{2}\right)}
 {2  s_{13}s_{24} s_{35} s_{14} s_{25}}
h_5^{\frac{d}{2}-r-1} P^{(d-2r-2)}
\end{eqnarray}
is invariant with respect to the change $d \rightarrow d \pm 2l $, where $l$ is
integer, so that this sum may be absorbed into the periodic constant that we
denoted by $\Pi_p(d)$.
Changing the summation index in the remaining sum as $r \rightarrow -r-1$, we
obtain Eq.~(\ref{solution_i5p}).
To obtain the solution of the difference equation~(\ref{equ_4_I5bar}) in terms
of convergent series, one may choose either Eq.~(\ref{solution_i5m}) or
Eq.~(\ref{solution_i5p}) depending on the kinematics.

The dependence of the arbitrary periodic functions $\Pi_m(d)$ and $\Pi_p(d)$
on the scalar invariants $s_{ij}$ may be constructed from a system of
differential equations which follows from the one for the integral
${I}^{(d)}_5$.
For the integral ${I}^{(d)}_5$, we derive a system consisting of 5 differential
equations of the form
\begin{eqnarray}
&&s_{13}s_{14}s_{24}s_{25}s_{35}~h_5~ s_{ij}
\frac{\partial {I}^{(d)}_5(s_{13},s_{14},s_{24},s_{25},s_{35})}
{\partial s_{ij}}
= (d-5) \sum_{k=1}^{5}J_k R^{(k)}_{ij}
\nonumber \\
&&~~~~~~~~~
+\left(\phi_{ij} -\frac{d}{2}s_{13}s_{14}s_{24}s_{25}s_{35} s_{ij} 
\frac{\partial h_5}{\partial s_{ij}}\right)
{I}^{(d)}_5(s_{13},s_{14},s_{24},s_{25},s_{35}), 
\label{dif_equ_for_i5}
\end{eqnarray}
where $(i,j)=(1,3),(1,4),(2,4),(2,5),(3,5)$ is not summed over, 
$J_k$ are the box integrals defined in Eq.~(\ref{I4integrals}), and
$R^{(k)}_{ij}$ and $\phi_{ij}$ are polynomials in $s_{kr}$. 
To derive this system of equations, we use the method proposed in
Ref.~\cite{Tarasov:1996bz}.
Some details of the derivation are presented in Appendix~B.
The derivation of such equations is done with the help of the computer program
package Maple.
Explicit expressions for $R^{(k)}_{ij}$ and $\phi_{ij}$ are given in
Eqs.~(\ref{R_polynoms}) and (\ref{phi_polynoms}) in Appendix~B, respectively.
Using Eq.~(\ref{i5_redefinition}), one may obtain from
Eq.~(\ref{dif_equ_for_i5}) a system of equations for the integral
$\overline{I}^{(d)}_5$.
Substituting Eq.~(\ref{solution_i5m}) into this system, we obtain the following
system of equations for the periodic function $\Pi_m(d)$ after a rather tedious
calculation:
\begin{eqnarray}
&&s_{13} \frac{\partial \Pi_m(d)}{\partial s_{13}}
=\frac{\phi_{13}}{h_5 s_{13}s_{14}s_{24}s_{25}s_{35}} \Pi_m(d),
~~~~~~~~~~~~~
s_{14} \frac{\partial \Pi_m(d)}{\partial s_{14}}
=\frac{\phi_{14}}{h_5 s_{13}s_{14} s_{24}s_{25}s_{35}} \Pi_m(d),
\nonumber \\
&&s_{24}\frac{\partial \Pi_m(d)}{\partial s_{24}}
=\frac{\phi_{24}}{h_5 s_{13}s_{14}s_{24} s_{25}s_{35}} \Pi_m(d),
~~~~~~~~~~~~~
s_{25} \frac{\partial \Pi_m(d)}{\partial s_{25}}
=\frac{\phi_{25}}{h_5 s_{13}s_{14} s_{24}s_{25}^2s_{35}} \Pi_m(d),
\nonumber \\
&&s_{35} \frac{\partial \Pi_m(d)}{\partial s_{35}}
=\frac{\phi_{35}}{h_5 s_{13}s_{14}s_{24} s_{25}s_{35}} \Pi_m(d).
\label{system_pm}
\end{eqnarray}
These differential equations do not depend explicitly on $d$ and are, thus,
much simpler than those for the integral $I^{(d)}_5$ itself.
The solution of this system of differential equation with respect to $s_{ij}$
for $\Pi_m(d)$ obtained with the help of the computer program package Maple
reads:
\begin{equation}
\Pi_m(d) =  \frac{\kappa(d)~h_5^{\frac52} }{(s_{13}s_{14}s_{35}s_{24}s_{25})^3},
\label{Pi_mom_dep}
\end{equation}
where $\kappa(d)$ is an arbitrary periodic constant,
\begin{equation}
\kappa(d+2) = \kappa(d),
\end{equation}
which is independent of the scalar invariants $s_{ij}$.
The system of differential equations for $\Pi_p(d)$ looks similar to
Eq.~(\ref{system_pm}).
An arbitrary periodic constant $\kappa(d)$ may be determined from ${I}^{(d)}_5$
calculated for some particular kinematics. 
Usually, setting some of the scalar invariants $s_{ij}$ to zero greatly
simplifies the computation of the integral.
But in our case, as may be seen from Eq.~(\ref{Pi_mom_dep}), $\kappa(d)$ cannot
be determined from $I_5^{(d)}$ calculated for such kinematics because the term
proportional to $\kappa(d)$ drops out.
For the same reason, one cannot use $I_5^{(d)}$ calculated for kinematics with
$h_5=0$.
Instead, we determine the periodic function $\Pi_{m}(d)$ by comparing the
limiting value of our analytic result for $|d|\rightarrow \infty$ with the
analogous value obtained from the integral representation of
Eq.~(\ref{I5_int_rep}) by exploiting the steepest-descent method described in
details in Ref.~\cite{Fleischer:2003rm}.

Without loss of generality, we henceforth assume the following hierarchy
between the scalar invariants:
\begin{equation}
-s_{13} > -s_{14},-s_{35} > -s_{25}>-s_{24}.
\label{hierarhija}
\end{equation}
For the case when the scalar invariants $s_{ij}$ satisfy the conditions of
Eq.~(\ref{hierarhija}) and additionally
\begin{equation}
-s_{14}-s_{35}+s_{13}>0,
\label{dop_condition}
\end{equation}
the integrals $I_4^{(d)}$ in Eq.~(\ref{I4integrals}) may be written as
\begin{eqnarray}
&&I_4^{(d)}(0,0,0,s_{25};s_{24},s_{35})
 =\chi^{(d)}(s_{25},s_{25},s_{35},s_{24})
 -{\eta}^{(d)}(s_{24},s_{25},s_{35},s_{24})
 -\chi^{(d)}(s_{35},s_{25},s_{35},s_{24}),
\nonumber \\
&&I_4^{(d)}(0,0,0,s_{13};s_{14},s_{35})
= \chi^{(d)}(s_{13},s_{13},s_{14},s_{35})
 -\chi^{(d)}(s_{35},s_{13},s_{14},s_{35})
 -\chi^{(d)}(s_{14},s_{13},s_{14},s_{35})
\nonumber \\
&&~~~~~~~~~~~~~~~~~~~~~~~~~~~~~~~~~~~~~~~~~~~~~~~~~~~~
-\frac{2 \pi (d-3)}{s_{35}s_{14} \sin \frac{\pi d}{2} }
~I_2^{(d)}\left( \frac{-s_{35} s_{14}}{s_{35}+s_{14}-s_{13}} \right),
\nonumber \\
&&I_4^{(d)}(0,0,0,s_{24};s_{25},s_{14})
={\eta}^{(d)}(s_{24},s_{24},s_{25},s_{14})
-\chi^{(d)}(s_{14},s_{24},s_{25},s_{14})
-\chi^{(d)}(s_{25},s_{24},s_{25},s_{14})
\nonumber \\
&&~~~~~~~~~~~~~~~~~~~~~~~~~~~~~~~~~~~~~~~~~  
-\frac{4 \pi (d-3)}{s_{14} s_{25}    \tan \frac{\pi d}{2}}
             ~I_2^{(d)}\left(\frac{s_{25} s_{14}}
             {s_{25} + s_{14} - s_{24}} \right) ,
\nonumber 
\end{eqnarray}
\begin{eqnarray}
&&
I_4^{(d)}(0,0,0,s_{35};s_{13},s_{25})
=\chi^{(d)}(s_{35},s_{35},s_{13},s_{25})
-{\eta}^{(d)}(s_{25},s_{35},s_{13},s_{25})
-\chi^{(d)}(s_{13},s_{35},s_{13},s_{25}),
\nonumber \\
&&
I_4^{(d)}(0,0,0,s_{14};s_{24},s_{13})
=\chi^{(d)}(s_{14},s_{14},s_{24},s_{13})-\chi^{(d)}(s_{13},s_{14},s_{24},s_{13})
\nonumber \\
&&~~~~~~~~~~~~~~~~~~~~~~~~~~~~~~~~~~~~~~~~~~~~~~~~~~~~~~~~~~
-{\eta}^{(d)}(s_{24},s_{14},s_{24},s_{13}),
\label{I4_in_terms_h_chi}
\end{eqnarray}
where
\begin{eqnarray}
&&{\eta}^{(d)}(s,s_{14},s_{24},s_{13})=
\frac{4(d-3)I_2^{(d)}(s)}{(d-4)s_{13}s_{24}}
       \Fh21\FSC{1,\frac{d}{2}-2}{\frac{d}{2}-1},
\nonumber \\
&& \nonumber \\
&& \chi^{(d)}(s,s_{14},s_{24},s_{13})=
\frac{- 4(d-3)I_2^{(d)}(s)}{(d-6) s (s_{13}+s_{24}-s_{14}) }
       \Fh21\FIR{1,3-\frac{d}{2}}{4-\frac{d}{2}},
\end{eqnarray}
and $I_2^{(d)}(p^2)$ is the one-loop massless propagator-type integral
\begin{equation}
I_2^{(d)}(p^2)=
\frac{1}{i \pi^{d/2}} \int \frac{d^dq}{q^2(q-p)^2}=
 \frac{-\pi^{\frac32}~(-p^2)^{\frac{d}{2}-2}}
{2^{d-3}  \Gamma\left(\frac{d-1}{2}\right)
 \sin \frac{\pi d }{2}}.
\label{I2_00p}
\end{equation}
An explicit derivation of these results using the method of dimensional
recurrences may be found in Ref.~\cite{Kniehl:2009b}. 
Results for these integrals were also obtained in Ref.~\cite{Bern:1992em}
using a different method.

To evaluate $I_5^{(d)}$, we use the solution in the form of
Eq.~(\ref{solution_i5m}).
This solution may be used if
\begin{equation}
\left|\frac{4}{h_5 s_{ij}}\right| \leq 1.
\label{region_I}
\end{equation}
As we shall see later, the quantities $4/(h_5s_{ij})$ emerge as expansion
parameters in the resulting hypergeometric series.
If condition~(\ref{region_I}) is fulfilled, then we may obtain a convergent
series using Eq.~(\ref{solution_i5m}).
If this  condition is not fulfilled, then we may use Eq.~(\ref{solution_i5p}).
In this case, the inverse quantities, $ h_5 s_{ij}/4 $, will be the expansion
parameters in the resulting hypergeometric series.

In the following, we obtain an analytic result assuming that
Eq.~(\ref{region_I}) is fulfilled for all scalar invariants $s_{ij}$. 
If this condition is not fulfilled for a particular term in $P^{(d)}$, then
an analytic result may be obtained by performing analytic continuations of  
the hypergeometric functions in the final result.
Another possibility to obtain the analytic result in this case is to repeat the
calculation using the solution of the form of Eq.~(\ref{solution_i5p}).
It should be noted that, if Eq.~(\ref{dop_condition}) is not satisfied, then
the arguments of all hypergeometric functions generated by the integral
$I_4^{(d)}(0,0,0,s_{13};s_{14},s_{35})$ exceed unity.
Analytic continuation of the result for this integral given in  
Eq.~(\ref{I4_in_terms_h_chi}) yields
\begin{equation}
I_4^{(d)}(0,0,0,s_{13};s_{14},s_{35})= \eta^{(d)}(s_{13},s_{13},s_{14},s_{35})
 -\eta^{(d)}(s_{35},s_{13},s_{14},s_{35})
 -\eta^{(d)}(s_{14},s_{13},s_{14},s_{35}).
\label{J2_dopcondi}
\end{equation}
Adopting Eq.~(\ref{region_I}), exploiting Eq.~(\ref{I4_in_terms_h_chi})
and, if $-s_{14}-s_{35}+s_{13}<0$, also Eq.~(\ref{J2_dopcondi}) for $J_2$, we
obtain a result in which each term is real.

The infinite sums resulting from Eq.~(\ref{solution_i5m}) may be written in
terms of known hypergeometric functions. 
As is evident from explicit expressions for the $I_4^{(d)}$ integrals, we must
compute three different types of sums.
The first one is related to the function $\chi^{(d)}$, 
\begin{eqnarray}
&&\sum_{r=0}^{\infty}\Gamma\left(\frac{d-2r-6}{2}\right)
h_5^{\frac{d}{2}-r-1}{\chi}^{(d-2r-2)}(s,a,b,c)
\nonumber \\
&&\nonumber \\
&&=\frac{-4h_5^{\frac{d}{2}-1}}{s~Q}\sum_{r=0}^{\infty}
I_2^{(d-2r-2)}(s)\frac{\Gamma\left(\frac{d}{2}-r-3\right)
}{h_5^r} \frac{(d-2r-5)}{(d-2r-8)}
\Fh21\Fx{1,r+4-\frac{d}{2}}{r+5-\frac{d}{2}}
\nonumber \\
&& =\frac{4(d-3)(d-5)\Gamma\left(\frac{d}{2}-4\right)}
{Q~ s^2 (x-1)}~h_5^{\frac{d}{2}-1}~I_2^{(d)}(s)~
F_3\left(1,1,\frac{7-d}{2},1, \frac{10-d}{2};
\frac{4}{s h_5} ,\frac{x}{x-1} \right),
\label{function_phi}
\end{eqnarray}
where
\begin{equation}
Q=c + b - a,\qquad
x=\frac{c b}{s Q}.
\end{equation}
The result of the summation of a term including the ${\eta }^{(d)}$ function
reads: 
\begin{eqnarray}
&&\sum_{r=0}^{\infty}\Gamma\left(\frac{d-2r-6}{2}\right)
h_5^{\frac{d}{2}-r-1} {\eta}^{(d-2r-2)}(s,a,b,c)
\nonumber \\
&&\nonumber \\
&&~~~~~~~~~~
=\frac{8(d-3)(d-5)}{b c s}
h_5^{\frac{d}{2}-1}\Gamma\left(\frac{d-6}{2}\right)
I_2^{(d)}(s)
\nonumber \\
&&~~~~~~~~~\times
\sum_{r = 0}^{\infty}
\frac{\left(\frac{7-d}{2}\right)_r}{\left(\frac{8-d}{2}\right)_r}
\left(\frac{4}{s~h_5}\right)^r \frac{1}{(d-2r-6)}
\Fh21\FSA{1,\frac{d-6}{2}-r}{\frac{d-4}{2}-r}.
\label{funchtion_H}
\end{eqnarray}
The simplest type of infinite series originates from the two terms in
Eq.~(\ref{I4_in_terms_h_chi}) without ${_2F_1}$ functions.
The contribution arising from the term proportional to $\varkappa_2$ reads:
\begin{eqnarray}
&&\sum_{r=0}^{\infty}\Gamma\left(\frac{d-2r-6}{2}\right)
~h_5^{\frac{d}{2}-r-1}~
\frac{(- 2 \pi) (d-2r-5)}{s_{35}s_{14} ~{\rm sin}\frac{\pi (d-2r-2)}{2}}
I_2^{(d-2r-2)}\left(\frac{-s_{35}s_{14}}{s_{35}+s_{14}-s_{13}}\right)
\nonumber \\
&&=\frac{8\pi(d-3)(d-5)}{(d-6) {\rm sin}\frac{\pi d}{2}}
\frac{(s_{14}+s_{35}-s_{13})}{s_{35}^2s_{14}^2}
I_2^{(d)}\left(\frac{-s_{35}s_{14}}{s_{35}+s_{13}-s_{13}}\right)
\nonumber \\
&& \times h_5^{\frac{d}{2}-1} \Gamma\left(\frac{d-4}{2}\right)
\Fh21\FIP{1,\frac{7-d}{2}}{\frac{8-d}{2}},
\label{contrib_kappa2}
\end{eqnarray}
and the one related to $\varkappa_3$ reads:
\begin{eqnarray}
&& \sum_{r=0}^{\infty}\Gamma\left(\frac{d-2r-6}{2}\right)
~h_5^{\frac{d}{2}-r-1}~\frac{(-4\pi)(d-2r-5)}{s_{14}s_{25}
\tan\frac{\pi (d-2r-2)}{2}}~I_2^{(d-2r-2)}\left(\frac{s_{14}s_{25}}
{s_{25}+s_{14}-s_{24}}\right) \nonumber \\
&& \nonumber \\
&& ~~=\frac{16 \pi (d-3)(d-5)}{(d-6)\tan\frac{\pi d}{2}}
\frac{(s_{14}+s_{25}-s_{24})}{s_{35}^2 s_{14}^2}
~h_5^{\frac{d}{2}-1} \Gamma\left(\frac{d-4}{2} \right)  
~I_2^{(d)}\left(\frac{s_{14}s_{25}}
{s_{25}+s_{14}-s_{24}}\right)
\nonumber \\
&&~~~~~~~\times
\Fh21\FDB{1,\frac{7-d}{2}}{\frac{8-d}{2}}.
\label{contrib_kappa3}
\end{eqnarray}

Inserting the sums of Eqs.~(\ref{function_phi}), (\ref{funchtion_H}),
(\ref{contrib_kappa2}), and (\ref{contrib_kappa3}) in Eq.~(\ref{solution_i5m}),
we obtain the following expression for the integral $I_5^{(d)}$:
\begin{eqnarray}
&&I_5^{(d)}(s_{13},s_{14},s_{24},s_{25},s_{35})
=\frac{-\pi^2~\Gamma \left(3-\frac{d}{2}\right)}{\sin^2\frac{\pi d}{2} }
~
\frac{ (-h_5)^{\frac{5-d}{2}}}{\sqrt{-s_{13}s_{14}s_{35}s_{24}s_{25}}}
\nonumber \\
&&~~~~~+
\frac{\varkappa_2 (s_{35}+s_{14}-s_{13})(d-8) \pi K }
{ 2 h_5 s_{35}^2 s_{14}^2 
~\sin \frac{\pi d}{2} }
  I_2^{(d)}\left(\frac{-s_{35} s_{14}}{s_{35}+s_{14}-s_{13}} \right)
 \Fh21\FIP{1,\frac{7-d}{2}}{\frac{8-d}{2}}
\nonumber \\
&& \nonumber \\
&&
+\frac{\varkappa_3 (s_{14}+s_{25}-s_{24})(d-8) \pi K}
{h_5 s_{14}^2 s_{25}^2  \tan \frac{\pi d}{2}}
 ~I_2^{(d)}\left(\frac{s_{14} s_{25} }{s_{14}+s_{25}-s_{24}}\right)
\Fh21\FDB{1,\frac{7-d}{2}}{\frac{8-d}{2}}
\nonumber \\
&& \nonumber \\
&&~~~~~~~~~~
  +\frac{(d-8)K}{h_5 s_{24}} I_2^{(d)}(s_{24})
  \left[\frac{\varkappa_1}{ s_{35} s_{24}} H^{(d)}\left(\frac{4}{h_5 s_{24}},
  \frac{(s_{35}+s_{24}-s_{25})}{s_{35}}\right)
\right.
\nonumber \\
&& \nonumber \\
&&\left.~~~~
  -\frac{\varkappa_3}{s_{14} s_{25}}
  H^{(d)}\left(\frac{4}{h_5 s_{24}},\frac{s_{24}(s_{14}+s_{25}-s_{24})}{s_{14}s_{25}}\right)
 +\frac{\varkappa_5}{s_{13} s_{24}}
  H^{(d)}\left(\frac{4}{h_5 s_{24}},\frac{s_{13}+s_{24}-s_{14}}{s_{13}}\right)
  \right]
\nonumber \\
&& \nonumber \\
&&~~~ -\frac{K}{h_5 s_{13}} I_2^{(d)}(s_{13})
  \left[\frac{\varkappa_2}{(s_{13}-s_{35})(s_{13}-s_{14})}
  \Phi^{(d)}\left( \frac{4}{h_5s_{13}},\frac{s_{35}s_{14}}{(s_{13}-s_{35})(s_{13}-s_{14})}
           \right)
\right.
\nonumber 
\\
&&\left.
~~~
  +\frac{\varkappa_4}{(s_{13}-s_{35})s_{13}}
  \Phi^{(d)}\left(\frac{4}{h_5 s_{13}},\frac{-s_{25}}{s_{13}-s_{35}} \right)
  +\frac{\varkappa_5}{(s_{13}-s_{14})s_{13}}
   \Phi^{(d)}\left(\frac{4}{h_5 s_{13}}, \frac{-s_{24}}{s_{13}-s_{14}}\right)
   \right]
\nonumber
\end{eqnarray}
\begin{eqnarray}
&&
 + \frac{K}{h_5 s_{14}} I_2^{(d)}(s_{14})\left[
 \frac{\varkappa_2}{(s_{13}-s_{14})s_{14}}
 \Phi^{(d)}\left( \frac{4}{h_5 s_{14}},\frac{s_{35}}{s_{13}-s_{14}}
 \right)
-\frac{\varkappa_3}{(s_{14}-s_{24}) s_{14}}
 \Phi^{(d)}\left(\frac{4}{h_5 s_{14}},\frac{-s_{25}}{s_{14}-s_{24}}
 \right)
\right.
\nonumber \\ 
&& \nonumber \\
&&~~~\left.
+\frac{\varkappa_5}{(s_{14}-s_{24})(s_{13}-s_{14})}
 \Phi^{(d)}\left( \frac{4}{h_5 s_{14}},
\frac{-s_{13} s_{24}}{(s_{14}-s_{24})(s_{13}-s_{14})}
 \right)
   \right]
\nonumber \\
&& \nonumber \\
&&
 +\frac{K}{h_5 s_{25}} I_2^{(d)}(s_{25})\left[
   \frac{\varkappa_1}{(s_{25}-s_{35}) (s_{24}-s_{25})}
  \Phi^{(d)}\left( \frac{4}{h_5 s_{25}},
\frac{-s_{24}s_{35}}{(s_{25}-s_{35})(s_{24}-s_{25})}
  \right)
\right.
\nonumber \\
&& \nonumber \\
&&~~~\left. 
 +\frac{\varkappa_4(d-8) }{s_{25} s_{13}}
   H^{(d)}\left(\frac{4}{h_5 s_{25}},\frac{s_{25}+s_{13}-s_{35}}{s_{13}} \right)
 +\frac{\varkappa_3}{(s_{24}-s_{25})s_{25}}
  \Phi^{(d)}\left(\frac{4}{h_5 s_{25}},
     \frac{s_{14}}{s_{24}-s_{25}}
  \right)\right]
\nonumber \\
&& \nonumber \\
&&
 +\frac{K}{h_5 s_{35}} I_2^{(d)}(s_{35})\left[
 \frac{ \varkappa_1}{ (s_{25}-s_{35}) s_{35}}
 \Phi^{(d)}\left(\frac{4}{h_5 s_{35}}, \frac{s_{24}}{s_{25}-s_{35}}
\right)
 +\frac{\varkappa_2}{ (s_{13}-s_{35})s_{35}}
 \Phi^{(d)}\left(\frac{4}{h_5 s_{35}}, \frac{s_{14}}{s_{13}-s_{35}}
\right)
\right.
\nonumber \\
&& \nonumber \\
&&~~~~~~~~~~~~~~~~\left. 
-\frac{\varkappa_4}{(s_{25}-s_{35}) (s_{13}-s_{35})}
  \Phi^{(d)}\left( \frac{4}{h_5 s_{35}},
\frac{s_{25} s_{13}}{(s_{25}-s_{35}) (s_{13}-s_{35})}
 \right)\right],
\label{i5_euclid}
\end{eqnarray}
where
\begin{equation}
K = \frac{-8(d-3)(d-5)}{(d-6)(d-8) s_{13} s_{14} s_{24} s_{25} s_{35}},
\end{equation}
\begin{eqnarray}
\label{Phi}
\Phi^{(d)}(w,z) &=& \sum_{r=0}^{\infty}
\frac{\left(\frac{7-d}{2}\right)_r}
     {\left(\frac{10-d}{2}\right)_r} w^r
\Fh21\Fe{1,1 }{r+5-\frac{d}{2}} 
\nonumber \\
&& \nonumber \\
&=& F_3\left(1,1,\frac{7-d}{2},1,\frac{10-d}{2}; w,z\right),
\\
H^{(d)}(w,z)&=&\sum_{r=0}^{\infty}\frac{\left(\frac{7-d}{2}\right)_r}
     {\left(\frac{8-d}{2}\right)_r}~\frac{ w^r}{(d-2r-6)}
~\Fh21\Fe{1,\frac{d-6}{2}-r}{\frac{d-4}{2}-r}
\nonumber \\
&& \nonumber \\
& =& \frac{1}{(d-6)}\sum_{l=0}^{\infty}z^l
 \frac{\left(\frac{d-6}{2}\right)_l}{\left(\frac{d-4}{2}\right)_l}
\Fh32\Fw{1,\frac{7-d}{2},\frac{6-d}{2}-l}
        {\frac{8-d}{2},\frac{8-d}{2}-l},
\label{H}
\end{eqnarray}
and the quantities $\varkappa_i$ are defined in Eq.~(\ref{P_ot_d}).
The functions $H^{(d)}$ and $\Phi^{(d)}$ are related as
\begin{equation}
H^{(d)}(w,z)=
\frac{1}{(1-z)(d-8)} \Phi^{(d)}\left(w,\frac{1}{1-z}\right)
-\frac{\pi (-z)^{3-\frac{d}{2}}}{2 \sin \frac{\pi d}{2}}
\Fh21\FwZ{1,\frac{7-d}{2}}{\frac{8-d}{2} }.
\end{equation}
The Appell function $F_3$ admits the following one-fold integral
representation: 
\begin{equation}
F_3\left(1,1,\frac{7-d}{2},1,\frac{10-d}{2};w,z\right)=
\frac{\Gamma\left(\frac{10-d}{2}\right)}
{\Gamma\left(\frac32\right)\Gamma\left(\frac{7-d}{2}\right)}
\frac{1}{\sqrt{z}} \int_0^1\frac{(1-u)^{\frac{5-d}{2}}}
{1-w+uw}\, \frac{\arcsin\sqrt{u z}}{\sqrt{1-uz }}du.
\label{onefoldintrepF3}
\end{equation}

It is interesting to note that changing the space-time dimension $d$ in 
Eq.~(\ref{onefoldintrepF3}) by one unit, $d \rightarrow d+1$, we obtain the
Appell function $F_3$, which we already encountered in the calculation of the
one-loop master integral entering the calculation of radiative corrections to
Bhabha scattering \cite{Kniehl:2009pv}.
As a matter of fact, we observed in Ref.~\cite{Kniehl:2009pv} that, at the
one-loop level, the set of hypergeometric functions appearing in the results
for $n$-point integrals with massive propagators also appear (up to the change
$d \rightarrow d+1$) in the calculation of ($n+1$)-point integrals with all
propagators being massless. 
For example, the one-loop propagator integral with two different masses is
expressible in terms of two Gauss hypergeometric functions 
${_2F_1(1,(d-1)/2;d/2;z)}$ with different arguments, while the result for the 
one-loop vertex integral with massless propagators is expressible in terms of
the ${_2F_1(1,(d-2)/2;(d-1)/2;z)}$ function with different arguments.
The result for the one-loop vertex integral with arbitrary masses and external
momenta \cite{Fleischer:2003rm} is expressible in terms of the Appell function
$F_1((d-2)/2,1,1/2,d/2;x,y)$ and the Gauss hypergeometric function
${}_2F_1(1,(d-2)/2;(d-1)/2;z)$ with different arguments,
while the result for the box integral with massless propagators and arbitrary
external momenta is expressible in terms of the same functions up to the shift 
$d\rightarrow d+1$, as was observed in Ref.~\cite{Kniehl:2009b}.
We stress that the number of terms with hypergeometric functions appearing in
the expressions for one-loop integrals with massive propagators and the
arguments of these hypergeometric functions are different from those for their
counterparts with massless propagators, but the sets of hypergeometric
functions are the same up to the shift in $d$ mentioned above.

The analytic continuation of the result in Eq.~(\ref{i5_euclid}) amounts to the
analytic continuation of the hypergeometric functions $F_3$ and ${_2F_1}$ and
the factors in front of these functions.
From Eq.~(\ref{i5_euclid}), we may obtain the value of the integral in any
region by using the usual $i\epsilon$ prescription and observing that
$I_5^{(d)}$ is manifestly real in the Euclidean region where all scalar
invariants $s_{ij}$ are negative.
The analytic continuation of the Gauss hypergeometric function ${_2F_1}$ is
well understood \cite{Erdely}.
Several useful formulae are given in Appendix~A.
The analytic continuation of the Appell function $F_3$ may be obtained from the
series representation of Eq.~(\ref{Phi}) by the analytic continuation of the
${_2F_1}$ function.
We notice that the well-known formula for the analytic continuation of the
$F_3$ function \cite{Erdely} in terms of the Appell function $F_2$ is not
applicable if both arguments of the $F_3$ function in Eq.~(\ref{Phi}) are
large.
In this case, one may proceed by analytically continuing the functions
${_2F_1}$ and ${_3F_2}$ in Eqs.~(\ref{Phi}) and (\ref{H}).
For example, the analytic continuation of the ${_3F_2}$ function in
Eq.~(\ref{H}) is
\begin{eqnarray}
&&\Fh32\Fw{1,\frac{7-d}{2},\frac{6-d}{2}-l}
 {4-\frac{d}{2},4-\frac{d}{2}-l}
=\frac{(6-d)(d-6+2l)}{w (d-5)(d-4+2l)}
     ~ \Fh32\Fiw{1,\frac{d-4}{2},\frac{d-4}{2}+l}
         {\frac{d-3}{2},\frac{d-2}{2} +l}
\nonumber \\
&& \nonumber \\
&&~~~~~~~~~~~~~~~~~~~
 +\frac{\Gamma\left(\frac{8-d}{2}\right)(d+2l-6) 
  \Gamma\left(\frac{d-5}{2}\right)}
 { \sqrt{\pi} ~(2l+1)} (-w)^{\frac{d-7}{2}}~ 
   \Fh21\Fiw{\frac12, l+\frac12}{l+\frac32}
\nonumber \\
&& \nonumber \\
&&~~~~~~~~~~~~~~~~~~~
 - \frac{\pi (d+2l-6)  \Gamma\left(\frac{8-d}{2}\right) 
   \Gamma\left(l + \frac12 \right)}
   {2~ {\rm  sin} \frac{\pi  d}{2}~ \Gamma\left( \frac{7-d}{2} \right) l!}
   (-1)^l (-w)^{\frac{d}{2} + l-3}.
\end{eqnarray}
Substituting this formula into Eq.~(\ref{H}), we obtain the following
representation for the $H^{(d)}(w,z)$ function for $|w|>1$:
\begin{eqnarray}
H^{(d)}(w,z) &=& \frac{(6-d)}{(d-5)(d-4) w}~
\phi^{(d-1)}\left(\frac{1}{w},z\right)
-\frac{\pi^{\frac32}~\Gamma\left(\frac{8-d}{2}\right)}
{2~ {\rm sin} \frac{\pi d}{2} ~\Gamma\left(\frac{7-d}{2}\right)}
~\frac{ (-w)^{\frac{d-6}{2}} }{\sqrt{1-w z}}
\nonumber \\
&& \nonumber \\
&-& \frac{(-w)^{\frac{d-6}{2}}}{ \sqrt{\pi ~( w z-1)}}
\Gamma\left(4-\frac{d}{2}\right) 
\Gamma\left(\frac{d-5}{2}\right)                
\arcsin \sqrt{\frac{(1-w z)}{w (1-z)} } ,
\label{H_w_large}
\end{eqnarray}
where
\begin{eqnarray}
\phi^{(d)}(x,y)&=&
F^{1;2;1}_{1;1;0} \left[^{\frac{d-3}{2}:~ \frac{d-3}{2},~1;~~~~ 1;}
_{\frac{d-1}{2}:~~~~~~ \frac{d-2}{2};~~-;}~~x,y\right]=
\sum_{r=0}^{\infty} \sum_{s=0}^{\infty} 
\frac{\left( \frac{d-3}{2} \right)_{r+s}}
{\left( \frac{d-1}{2} \right)_{r+s}}~\frac{\left( \frac{d-3}{2} \right)_r}
{\left( \frac{d-2}{2} \right)_r} ~x^r y^s
\nonumber \\
&& \nonumber \\
&=&\sum_{r=0}^{\infty}
 \frac{\left(\frac{d-3}{2}\right)_r \left(\frac{d-3}{2}\right)_r }
      {\left(\frac{d-2}{2}\right)_r \left(\frac{d-1}{2}\right)_r }
      ~x^r
  \Fh21\Fy{1,\frac{d-3}{2}+r}{\frac{d-1}{2}+r},
\label{fii}
\end{eqnarray}
and $F^{1;2;1}_{1;1;0} $ is the Kamp\'e de F\'eriet function \cite{ApKdF}.
To analytically continue the $H^{(d)}(w,z)$ function into the region $w,z>1$,
one may use Eq.~(\ref{H_w_large}).
The analytic continuation of the $\phi^{(d)}(w,z)$ function into the region
$|z|>1$ may be obtained by the analytic continuation of the ${_2F_1}$ function
in Eq.~(\ref{fii}) using Eq.~(\ref{degenerated2F1_z1}) from Appendix~A. 

There are several relations for the $F_3$ function which may be useful for its
analytic continuation and performing its $\varepsilon$ expansion.
We present here two formulae for the analytic continuation of the $F_3$
function with large first argument $x$ and $y<0$.
One such relation follows from the results given in Ref.~\cite{Kniehl:2009pv}
and reads:
\begin{eqnarray}
&&F_3\left(1,1,\frac{7-d}{2},1,\frac{10-d}{2},x,y\right)
= \frac{(d-6)(d-8)}{(d-3)(d-5) (x y-y-x) x}
\left\{ \Fh21\FBX{1,\frac12}{\frac{d-1}{2}}
\right.
\nonumber \\
&&~~~~\left.
+\frac{(d-4)}{x(d-1)}
F_3\left(\frac12,1,1,\frac{d-2}{2},\frac{d+1}{2},
\frac{y}{x+y-x y} ,\frac{1}{x}\right)
\right\}
\nonumber \\ 
&&~~~~
+\frac{2}{\sqrt{\pi}}
\Gamma\left(\frac{d-5}{2} \right) \Gamma\left(\frac{10-d}{2} \right)
  \sqrt{1-x} ~ (-x)^{\frac{d-8}{2}}
\Fh21\FXM{1, 1}{\frac32 }
\nonumber \\
&& \nonumber \\
&&~~~~
- \frac{(d-8)}{ (d-5) ~x(1- y)}
\Fh21\FYM{1, \frac{d-4}{2} }{\frac{d-3}{2}} 
\Fh21\FYC{1,3-\frac{d}{2} }{4-\frac{d}{2} }. 
\end{eqnarray}
The $F_3$ function with appropriate parameters admits the one-fold
integral representation
\begin{equation}
F_3\left(\frac12,1,1,\frac{d-2}{2},\frac{d+1}{2};
~w,z\right)=
\frac{\Gamma\left(\frac{d+1}{2}\right)}
{\sqrt{\pi}~\Gamma\left(\frac{d-2}{2}\right)}\frac{1}{w}
\int_0^1 \frac{(1-v)^{\frac{d-4}{2}}}{1-z+vz}
~\ln \frac{1+\sqrt{w v  }}
        {1-\sqrt{w  v }}~d v,
\label{secondF3}
\end{equation}
which may also be used for the analytic continuation.
Another formula for the analytic continuation into the region of large $x$
and $y<0$ connects the $F_3$ function and the Horn function $H_2$,
\begin{eqnarray}
&&F_3\left(1,1,\frac{7-d}{2},1,\frac{10-d}{2};x,y \right)
=\frac{(8-d)}{x(d-5)}H_2\left(\frac{d-6}{2},1,1,1,\frac{d-3}{2};
\frac{1}{x},-y\right)
\nonumber \\
&& \nonumber \\
&&~~~~~~~~~~~~~~~ +\frac{2}{\sqrt{\pi}}
 \Gamma\left(\frac{d-5}{2}\right) \Gamma\left(\frac{10-d}{2}\right)
(-x)^{\frac{d-8}{2}} \sqrt{1-x}
\Fh21\FXM{1, 1}{\frac32 }.
\end{eqnarray}

We checked Eq.~(\ref{i5_euclid}) by comparing its numerical values calculated
for some specific values of $d$ and $s_{ij}$ satisfying the condition of
Eq.~(\ref{hierarhija}) with the result of the direct numerical evaluation of
the four-fold integral representation given by Eq.~(\ref{I5_int_rep}) in
Appendix~A using the computer algebra program package Maple.
In all cases, we found perfect agreement to all valid digits of the numerical
results (usually about 12).
For example, we performed a numerical calculation for $d=21/2$ and
$d=10-2 \varepsilon$, where $\varepsilon=1/10000$, setting
$s_{13}=-1000+\delta$, $s_{14}=-65+\delta$, $s_{35}=-190+\delta$,
$s_{24}=-1/20+\delta$, and $s_{25}=-1/10+\delta$ with $\delta=i 10^{-20}$.

\section{Case of one vanishing variable}

In practical applications, such as the analytic continuation of the
pentagon integral with the help of functional equations
\cite{Kniehl:2009pv,Tarasov:2008hw}, we need the value of the integral
$I_5^{(d)}$ with one or more vanishing kinematic variables.
As was shown in Ref.~\cite{Tarasov:2008hw}, the one-loop vertex integral
with arbitrary masses and external momenta may be expressed in terms of a
vertex integral, in which two masses and the square of one external momentum
vanish, with the help of functional equations.
The formulae for the analytic continuation of the vertex integral with
arbitrary masses contain integrals with all masses vanishing. 
The relation between the master integrals of Bhabha scattering and heavy-quark
production derived in Ref.~\cite{Kniehl:2009pv} includes an integral
with massless propagators.
In all these cases, relations connecting integrals with different kinematical
variables include integrals with simpler kinematics, in which either some
masses or some squared momenta are equal to zero.
We observed a rather similar situation for the pentagon integral.
Functional equations relevant for the analytic continuation of a considered
integral include the value of this integral with one scalar invariant taken to
be zero. 
For this reason, we wish to present here the value of the integral $I_5^{(d)}$
with one scalar invariant taken to be zero.
Due to the symmetry~(\ref{simmetry}) of this integral, it is sufficient to
consider the case when $s_{24}=0$.
A detailed discussion of functional equations for pentagon-type integrals
will be presented in a separate publication \cite{Kniehl:2009b}.

For $s_{24}=0$, we obtain the following result:
\begin{eqnarray}
&&
I_5^{(d)}(s_{13},s_{14},0,s_{25},s_{35})\nonumber \\
&& \nonumber \\
&&
=\frac{-8(d-3)(d-5)}{(d-6)(d-8)(s_{14} s_{35}-s_{14}s_{25}+s_{13}s_{25})}
\left\{
\frac{\pi (d-8) (s_{14}+s_{25})}{s_{14} s_{25} \tan\frac{\pi d}{2}}
I_2^{(d)}\left(\frac{s_{14} s_{25}}{s_{14}+s_{25}}\right)
\right.
\nonumber \\
&& \nonumber \\
&&
+\frac{s_{35}s_{14}}{(s_{13}-s_{35})(s_{13}-s_{14})s_{13}} I_2^{(d)}(s_{13})
\Fh21\FZA{1, 1}{5-\frac{d}{2}}
\nonumber \\
&& \nonumber \\
&&
+\frac{s_{25}}{s_{13}(s_{13}-s_{35})} I_2^{(d)}(s_{13})
\Fh21\FZB{1, 1}{5-\frac{d}{2}}
\nonumber \\
&& \nonumber \\
&&
-\frac{s_{35}}{s_{14}(s_{13}-s_{14})} I_2^{(d)}(s_{14})
\Fh21\FZC{1, 1}{5-\frac{d}{2}}
-\frac{ s_{25}}{s_{14}^2} I_2^{(d)}(s_{14})
 \Fh21\FZD{1, 1}{5-\frac{d}{2}} 
\nonumber \\
&& \nonumber \\
&&
-\frac{s_{14}}{s_{25}^2} I_2^{(d)}(s_{25})
\Fh21\FZE{1, 1}{5-\frac{d}{2}}
-\frac{(d-8)}{ (d-6) s_{25}} I_2^{(d)}(s_{25})
\Fh21\FZF{1, \frac{d}{2}-3}{\frac{d}{2}-2} 
\nonumber \\
&& \nonumber \\
&&
-\frac{s_{14}}{s_{35}(s_{13}-s_{35})} I_2^{(d)}(s_{35})
\Fh21\FZG{1, 1}{5-\frac{d}{2}} 
\nonumber \\
&& \nonumber \\
&&
+\frac{s_{13} s_{25}}{s_{35} (s_{25}-s_{35}) (s_{13}-s_{35})} I_2^{(d)}(s_{35})
\Fh21\FZH{1, 1}{5-\frac{d}{2}}
\nonumber \\
&& \nonumber \\
&&\left.
+\frac{\pi (s_{13}-s_{35}-s_{14}) (d-8)}{2 {\rm sin}\frac{\pi d}{2} ~ s_{14} s_{35}}
I_2^{(d)}\left(\frac{s_{35}s_{14}}{s_{13}-s_{35}-s_{14}}\right)
\right\}.
\end{eqnarray}
We observe that this result is significantly simplified compared to
Eq.~(\ref{i5_euclid}).
In fact, only ${_2F_1}$ functions remain.

\boldmath
\section{Asymptotic values of the integral $I_5^{(d)}$ for 
$|s_{13}|\rightarrow \infty$}
\unboldmath

If the scalar invariants fulfill the hierarchy of Eq.~(\ref{hierarhija}) with
$|s_{13}|$ being much larger than the other scalar invariants and
$s_{24} \neq s_{25}$, then we have
\begin{eqnarray}
h_5 &\cong&
\frac{(s_{24}-s_{25})^2 }{s_{24} s_{35} s_{14} s_{25}} s_{13}
\nonumber \\
&&{}
+\frac{2(s_{14} s_{24} s_{35}-s_{14} s_{25}^2-s_{24}^2 s_{35}
 +s_{24} s_{35} s_{25}+s_{14} s_{25} s_{35} + s_{14} s_{24} s_{25})}
 { s_{24} s_{35} s_{14} s_{25}} + O\left( \frac{1}{s_{13}}\right).
\end{eqnarray}
As one can see from the explicit expression of Eq.~(\ref{i5_euclid}), the
first argument of the two functions $\Phi^{(d)}$ and $H^{(d)}$ is always
proportional to $1/|s_{13}|$, so that it is sufficient to keep only the first
terms of the series in Eqs.~(\ref{Phi}) and (\ref{H}) in order to find the
asymptotically leading term of the integral $I_5^{(d)}$.
In $d=4-2 \varepsilon$ dimensions, this value reads:
\begin{eqnarray}
\lefteqn{\left.I_5^{(4 - 2\varepsilon )}(s_{13},s_{14},s_{24},s_{25},s_{35}) 
  \right|_{|s_{13}|\rightarrow \infty}}
\nonumber\\
&=&\frac{ \pi (1-2\varepsilon)}{ \sin \pi \varepsilon}
\frac{(s_{25}+s_{24})} {s_{14} s_{35}s_{24}s_{25}  }
~ I_2^{(4 - 2\varepsilon)}\left(\frac{s_{35} s_{14}}{ s_{13}}\right)
\left\{1-
\frac{(s_{25}-s_{24})^2}{s_{25}(s_{25}+s_{24})}
\Fh21\FCP{1,1-\varepsilon}{1+\varepsilon}
 \right\}
\nonumber \\
&&{}-\frac{\pi^2 \Gamma(1+\varepsilon) }{{\rm sin}^2 \pi \varepsilon}
\frac{(-s_{13})^{\varepsilon}[(s_{24}-s_{25})^2]^{\frac12 + \varepsilon}}
{(s_{14}s_{35}s_{24}s_{25})^{1+\varepsilon}}
+O\left(\frac{1}{|s_{13}|^{1-\varepsilon}}\right).
\end{eqnarray}

In $d=2-2\varepsilon$ dimensions, the result is
\begin{eqnarray}
\lefteqn{\left.I_5^{(2 - 2\varepsilon )}(s_{13},s_{14},s_{24},s_{25},s_{35}) 
\right|_{|s_{13}| \rightarrow \infty}
=\frac{4 \pi (1-4\varepsilon^2)(3+2\varepsilon)(s_{25}+s_{24}) s_{13}}
{s_{14}^2 s_{35}^2 (s_{24}-s_{25})^2 (2+\varepsilon)  \sin \pi \varepsilon }}
\nonumber \\
&&{}\times I_2^{(4 - 2\varepsilon)}\left(\frac{s_{35} s_{14}}{ s_{13}}\right)
\Fh21\FSD{1, \frac52+ \varepsilon }{ 3+ \varepsilon  } 
\nonumber \\
&&{}-\left(\frac{-s_{13} (s_{24}-s_{25})^2}
{s_{14}s_{35} s_{24} s_{25}}\right)^{\frac32+\varepsilon}
\frac{\pi^2 \Gamma(2+\varepsilon)}{ 
\sqrt{-s_{13} s_{14}s_{24}s_{25}s_{35}} ~{\rm sin}^2\pi \varepsilon }
+O\left(\frac{1}{|s_{13}|^{-\varepsilon}}\right).
\end{eqnarray}
We observe that, when one of the scalar invariants is large compared to the
others, the hypergeometric functions of two variables collapse to
hypergeometric function of just one variable.

In $d=6-2\varepsilon$ dimensions, the leading large-$|s_{13}|$ term of the
integral $I_5^{(d)}$ is somewhat more complicated, being
\begin{eqnarray}
\lefteqn{\left. I_5^{(6-2\varepsilon)}(s_{13},s_{14},s_{24},s_{25},s_{35})
\right|_{|s_{13}| \rightarrow \infty}=
-\frac{\pi^2 \Gamma(\varepsilon)}{ \sin^2 \pi \varepsilon}
\frac{\left[ \frac{-s_{13}(s_{25}-s_{24})^2}{s_{14}s_{35}s_{24}s_{25}} \right]^{\varepsilon}}
{s_{13}(s_{25}-s_{24})}}
\nonumber \\
&&{}-\frac{(1-2 \varepsilon) }{\varepsilon~ s_{13} (s_{25}-s_{24} )}
\nonumber \\
&&{}\times  \left \{
\frac{ - \pi(s_{24}+s_{25})}{(s_{25}-s_{24})  \sin \pi \varepsilon }
I_2^{(4-2\varepsilon)}\left(\frac{s_{35} s_{14}}{ s_{13}}\right)
\Fh21\FSD{1,  \frac12+\varepsilon  }{ 1+ \varepsilon } 
\right.
\nonumber \\
&&{}+\frac{2\pi}{\tan \pi \varepsilon } ~I_2^{(4-2\varepsilon)}
\left(\frac{s_{14}s_{25}}{s_{14}+s_{25}-s_{24}}\right)
+\frac{s_{24}}
{(1+ \varepsilon) ~ s_{14} }
 I_2^{(4-2\varepsilon)}(s_{14}) \Fh21\FPG{1, 1+\varepsilon}{2+\varepsilon}
\nonumber \\
&&{}-\frac{ s_{25}}{  (1+\varepsilon)  (s_{14}+s_{25}-s_{24})}
I_2^{(4-2\varepsilon)}(s_{14})
\Fh21\FPC{1, 1+\varepsilon}{2+\varepsilon}
\nonumber \\
&&{}-\frac{1}{\varepsilon }~
 I_2^{(4-2\varepsilon)}(s_{24})
 \Fh21\FSa{1, -\varepsilon}{1-\varepsilon}
\nonumber \\
&&{}-\frac{1}{ \varepsilon}
I_2^{(4-2\varepsilon)}(s_{24}) \Fh21\FSd{1, -\varepsilon}{1-\varepsilon}
+ I_2^{(4-2\varepsilon)}(s_{24})  \ln \frac{s_{14}-s_{24}}{s_{13}}
\nonumber \\
&&{}
-I_2^{(4-2\varepsilon)}(s_{25})  \ln \frac{s_{35}-s_{25}}{s_{13}}
 +\left[ I_2^{(4-2\varepsilon)}(s_{24}) - I_2^{(4-2\varepsilon)}(s_{25}) \right]
 \left[\gamma + \psi(-\varepsilon)\right]
\nonumber \\
&&{}-\frac{ s_{14} }
{ (1+\varepsilon)  (s_{14}+s_{25}-s_{24})}
I_2^{(4-2\varepsilon)}(s_{25})
\Fh21\FPH{1, 1+\varepsilon}{2+\varepsilon}
\nonumber \\
&&{}-\frac{s_{35} s_{24}}
{ (1+\varepsilon)  s_{25}
(s_{35}+s_{24}-s_{25})}
I_2^{(4-2\varepsilon)}(s_{25})
\Fh21\FPF{1, 1+\varepsilon}{2+\varepsilon}
\nonumber \\
&&{}+\frac{s_{24}}{  (1+\varepsilon)
~ (s_{35}+s_{24}-s_{25})}
I_2^{(4-2\varepsilon)}(s_{35}) \Fh21\FPD{1,1+\varepsilon}{2+\varepsilon} 
\nonumber \\
&&{}-\left.
\frac{s_{25}}{ (1+\varepsilon)
  ~s_{35} }
 I_2^{(4-2\varepsilon)}(s_{35})
 \Fh21\FPE{1, 1+\varepsilon}{2+\varepsilon}
\right\}
 +{\cal O}\left(\frac{1}{|s_{13}|^{2-\varepsilon}}\right),
\end{eqnarray}
where $\psi(x)$ is the logarithmic derivative of the Euler $\Gamma$ function,
\begin{equation}
\psi(x) = \frac{\Gamma'(x)}{\Gamma(x)},
\end{equation}
and $\gamma$ is the Euler-Mascheroni constant.

\section{Limit of multi-Regge kinematics}

In this section, we consider the pentagon integral $I_5^{(d)}$ with the
somewhat peculiar kinematics,
\begin{equation}
-s_{13} \gg -s_{14},-s_{35} \gg -s_{25},-s_{24}.
\label{MRK}
\end{equation}
In addition, we assume that the following relation among the scalar
invariants $s_{ij}$ holds:
\begin{equation}
s_{14}s_{35} \thicksim s_{13}s_{24} \thicksim s_{13}s_{25}. 
\end{equation}
Equivalently, one may define such an ordering by introducing the scaling
parameter $\lambda$, putting
\begin{equation}
s_{13}\rightarrow s_{13},
~~~s_{14} \rightarrow  \lambda s_{14},
~~~s_{35} \rightarrow  \lambda s_{35},
~~~s_{25} \rightarrow  \lambda^2 s_{25},
~~~s_{24} \rightarrow  \lambda^2 s_{24},
\label{lambda_arr}
\end{equation}
and taking the limit $\lambda \rightarrow 0$.
In fact, this arrangement corresponds to the multi-Regge kinematics.
Recently, the pentagon integral in $d=6-2\varepsilon$ dimensions in the
multi-Regge kinematics defined by Eq.~(\ref{lambda_arr}) was considered in
Ref.~\cite{DelDuca:2009ac}. 
The integral $I_5^{(d)}$ in $d=6-2\varepsilon$ dimensions is needed to
determine the $\varepsilon$ expansion of the integral $I_5^{(d)}$ in
$d=4 - 2\varepsilon$ dimensions.

Here, we consider the integral $I_5^{(d)}$ in the multi-Regge-kinematics limit
of Eq.~(\ref{MRK}) directly in $d=4-2\varepsilon$ dimensions. 
To obtain the leading contribution in this limit, we perform the scaling of
Eq.~(\ref{lambda_arr}) and retain the leading terms in the limit
$\lambda \rightarrow 0 $.
The most divergent terms in Eq.~(\ref{i5_euclid}) are of the order
$\ln \lambda /\lambda^{4+2\varepsilon}$.
As will be seen later, such terms cancel out in the sum, so that the leading
asymptotic terms are of order $1/\lambda^{4+2\varepsilon}$.
In order to find the leading terms in the limit $\lambda \rightarrow 0$,
we must know the asymptotic behavior of the $H^{(d)}(w,z)$ function for
$z\rightarrow 1$ and that of the $\Phi^{(d)}(w,z)$ function for
$z \rightarrow  \infty$.
The leading and subleading terms of these functions in the respective limits
may be obtained from the series representations of Eqs.~(\ref{H}) and
(\ref{Phi}) by retaining the first leading terms of the expansions of the
${_2F_1}$ functions.
With the help of Eq.~(\ref{degenerated2F1_z1}) in Appendix~A, we obtain
\begin{eqnarray}
&&\left. H^{(d)}(w,z)\right|_{z\rightarrow 1}=
\frac12\left\{\frac{\pi}{\tan \frac{\pi d}{2}}
- \gamma -\psi\left(4-\frac{d}{2}\right)-\ln(1-z)\right\}
\Fh21\Fw{1,\frac{7-d}{2}}{\frac{8-d}{2}}
\nonumber \\
&&~~~~~~~~~~~~~~~~~~~~~~ -
\left. \frac12 \frac{\partial}{\partial \alpha}
\Fh32\Fw{1,\frac{7-d}{2},4-\frac{d}{2}+\alpha}
{4-\frac{d}{2},4-\frac{d}{2}}\right|_{\alpha=0} +O((1-z)\ln(1-z)).
\label{H_z_1}
\end{eqnarray}
In a similar way, the asymptotic behavior of the $\Phi^{(d)}$ function
for $z\rightarrow  \infty$ may be derived.
Keeping the logarithmic and constant terms of Eq.~(\ref{degenerated2F1_zinf})
in Appendix~A, we arrive at the following result:
\begin{eqnarray}
&&\left. \Phi^{(d)}(w,z)\right|_{z\rightarrow \infty}=
\frac{8-d}{2z}\left\{ 
\left[\gamma-\ln(-z)+
\psi\left(4-\frac{d}{2} \right)\right] \Fh21\Fw{1,\frac{7-d}{2}}{\frac{8-d}{2}}
\right.
\nonumber \\
&& \nonumber \\
&&\left.~~~~~~~~~~~~~~~~~~~~~~~~
+\left.
\frac{\partial}{\partial \alpha}
\Fh32\Fw{1,\frac{7-d}{2},4-\frac{d}{2}+\alpha}
{4-\frac{d}{2},4-\frac{d}{2}}\right|_{\alpha=0}
\right\} +O\left(\frac{\ln z}{z}\right).
\label{Phi_z_large}
\end{eqnarray}
Using  Eqs.~(\ref{H_z_1}) and (\ref{Phi_z_large}), and retaining only terms
contributing at orders $\ln \lambda/\lambda^{4+2\varepsilon}$ and
$1/\lambda^{4+2\varepsilon}$, we find from Eq.~(\ref{i5_euclid}) in the 
limit of multi-Regge kinematics the leading term,
\begin{eqnarray}
&&I_5^{(4-2\varepsilon)}(s_{13},s_{14},s_{24},s_{25},s_{35})
= \frac{-\pi^2 \Gamma(1+\varepsilon) (-\overline{h}_5)^{\frac12+\varepsilon}}
 {{\rm sin}^2\pi \varepsilon~\sqrt{-s_{13} s_{14} s_{24} s_{25} s_{35}}}
-\frac{2(1-4 \varepsilon^2)}
{(1+\varepsilon)  s_{13} s_{14}s_{35} s_{25}s_{24} \overline{h}_5 }
\nonumber \\
&& \nonumber \\
&&
\times
\left\{ 
 \frac{2(s_{13}s_{24}-s_{14}s_{35}-s_{13}s_{25})}{s_{24}}
 I_2^{(4-2\varepsilon)}(s_{24})
 H^{(4-2 \varepsilon)}\left(\frac{4}{ s_{24} \overline{h}_5 },
 \frac{s_{24}}{s_{25}} \right)
\right.
\nonumber \\
&& \nonumber \\
&&~~~~~
-
\frac{(s_{14}s_{35}+s_{13}s_{24}-s_{13}s_{25})}{s_{24}}
 I_2^{(4-2\varepsilon)}(s_{24}) \ln Y
 \Fh21\FhC{1, \frac32+\varepsilon}{2+\varepsilon}
 \nonumber \\
&& \nonumber \\
&&~~~~~
 +\frac{s_{24} (s_{14} s_{35}+s_{13} s_{24}-s_{13} s_{25})}
 {(2+\varepsilon) (s_{24}-s_{25}) s_{25}}
 I_2^{(4-2\varepsilon)}(s_{25}) 
 \Phi^{(4-2 \varepsilon)}\left(\frac{4}{s_{25} \overline{h}_5 },
 \frac{s_{24}}{s_{24} -s_{25}} \right)
\nonumber \\
&& \nonumber \\
&&~~~~~
 + \frac{(s_{13}s_{24}-s_{14}s_{35}-s_{13}s_{25})}{s_{25}}
  I_2^{(4-2\varepsilon)}(s_{25})
  \left( \ln Y + \frac{\pi} 
  {\tan\pi \varepsilon }
  \right)
 \Fh21\FhP{1, \frac32+\varepsilon}{2+\varepsilon}
\nonumber \\
&& \nonumber \\
&&\left.
 -\frac{ \pi s_{13}(s_{14}s_{35}+s_{13} s_{24}+s_{13} s_{25})}
 {{\rm sin} \pi \varepsilon~ s_{14}s_{35}}
  I_2^{(4-2 \varepsilon)}\left(\frac{s_{35}s_{14}}{s_{13}}\right) 
 \Fh21\FBR{1, \frac32+\varepsilon}{2+\varepsilon}
\right\},
\label{I5_MRK}
\end{eqnarray}
where
\begin{eqnarray}
Y&=&\frac{s_{13}(s_{25}-s_{24})}{s_{14}s_{35}},
\nonumber\\
\overline{h}_5&=&\frac{s_{14}^2 s_{35}^2+s_{13}^2 s_{24}^2
+s_{13}^2 s_{25}^2-2 s_{24} s_{13}^2 s_{25}
+2 s_{13} s_{35}s_{14}s_{25}+2s_{13}s_{24}s_{35}s_{14}}
{s_{13}s_{24}s_{35}s_{14}s_{25}}.
 \end{eqnarray}
As mentioned above, the terms with $\ln \lambda$ cancel, as well as
those involving the function $\psi(x)$, $\gamma$ and the derivative of 
the ${_3F_2}$ function with respect to a parameter.

The value of the integral $I_5^{(d)}$ for $d=6-2\varepsilon$ in the limit of
multi-Regge kinematics may be derived from Eq.~(\ref{dim_rec_i5}) using
Eq.~(\ref{I5_MRK}) and the asymptotic values of integrals $I_4^{(d)}$ given in
Eq.~(\ref{I4_in_terms_h_chi}).
We just note that it is again expressed in terms of $F_3$ and ${_2F_1}$
functions because the integrals $I_4^{(d)}$ add only ${_2F_1}$ functions.
An analytic expression for the integral $I_5^{(d)}$ in $d=6-2\varepsilon$
dimensions was recently obtained in Ref.~\cite{DelDuca:2009ac}.
The result of Ref.~\cite{DelDuca:2009ac} is given in terms of derivatives of
the Kamp\'e de F\'eriet function \cite{ApKdF} with respect to a parameter.
For a direct comparison of our result with that of Ref.~\cite{DelDuca:2009ac},
one needs an expression of the $F_3$ function in terms of derivatives of the
Kamp\'e de F{\'e}riet function, which, to our knowledge, is not currently
available.

\section{Conclusions}

In this paper, we evaluated the one-loop scalar pentagon integral in arbitrary
space-time dimension $d$ with on-shell external legs, massless internal lines,
and otherwise arbitrary scalar invariants.
Exploiting the method of dimensional recurrences, we obtained a result in terms
of the hypergeometric functions $F_3$ and ${_2F_1}$.
In our case, both functions admit one-fold integral representations suitable
for $\varepsilon$ expansions. 
Using the methods of Ref.~\cite{DelDuca:2009ac}, the on-shell pentagon integral
may be represented in terms of four-fold hypergeometric series, while the
method of dimensional recurrences advocated here just yields two-fold series.
 
The method of dimensional recurrences may also be applied to the evaluation 
 of the hexagon integral, which is needed for the calculation of the
$\cal{O}(\varepsilon)$ contributions to one-loop maximally-helicity-violating
amplitudes at one loop in ${\cal{N}}=4$ SYM theory.
To simplify derivations in this case, one may start from the dimensional
recurrences written in the limit of multi-Regge kinematics.
In our opinion, the method of dimensional recurrences is quite efficient to go
beyond the ``box approximation'' for the $n$-point one-loop integrals.

The $\varepsilon$ expansion of our results in Eqs.~(\ref{i5_euclid}) and
(\ref{I5_MRK}) will be presented in a future publication \cite{Kniehl:2009b}.

We expect that the results presented in this paper and our forthcoming one
\cite{Kniehl:2009b} may be conveniently incorporated in program packages for
the automated analytic computation of one-loop integrals in massless theories,
similar to the package for numerical calculations presented in
Ref.~\cite{vanHameren:2005ed}.

\section{Acknowledgments}

This work was supported in part by the German Federal Ministry for Education
and Research BMBF through Grant No.\ 05H09GUE, by the German Research
Foundation DFG through Grant No.\ KN~365/3--2, and by the Helmholtz Association
HGF through Grant No.\ HA~101.

\section{Appendix A}

\boldmath
\subsection{Integral representation of $I_5^{(d)}$}
\unboldmath

For numerical checks of the result for the integral
$I_5^{(d)}(s_{13},s_{14},s_{24},s_{25},s_{35})$ in Eq.~(\ref{i5_euclid}), we
use the Feynman parameterization
\begin{equation}
I_5^{(d)}(s_{13},s_{14},s_{24},s_{25},s_{35}) = - \Gamma\left(5-\frac{d}{2}\right)
\int_0^1...\int_0^1 dx_1 dx_2 dx_3 dx_4 
~ x_1^3 x_2^2 x_3 ~H_5^{\frac{d}{2}-5},
\label{I5_int_rep}
\end{equation}
where
\begin{eqnarray}
&&H_5=x_1x_2\left[
(1-x_3) (1-x_1) s_{13} + x_1 x_3 x_4 (1-x_2) s_{25}
+x_3 x_4 (1-x_1) s_{35}\right.
\nonumber \\
&&\left.
~~~~~~~ +x_1 x_3 (1-x_4) (1-x_2) s_{24} + x_1 x_2 x_3 (1-x_4) (1-x_3) s_{14}
\right].
\end{eqnarray}

\boldmath
\subsection{Useful formulae for the Gauss hypergeometric function ${_2F_1}$}
\unboldmath

The Gauss hypergeometric function ${_2F_1}$ has the following integral
representation:
\begin{eqnarray}
&&\Fh21\Fx{\alpha,\beta}{\gamma} =
\frac{\Gamma(\gamma)}{\Gamma(\beta)\Gamma(\gamma-\beta)}
~~ \int_0^1 du  \, u^{\beta-1}
(1-u)^{\gamma-\beta-1} (1-u x)^{-\alpha},
\nonumber \\
&& \nonumber \\
&&
 ~~~~~~~~~~~~~~~~~~~~~~~~~~~~~~~~~~~~~~~~~~~~~~~
{\rm Re}(\beta)>0, \quad {\rm Re}(\gamma-\beta) >0.
\end{eqnarray}

The following formulae are useful for the analytic continuation of the
${_2F_1}$ function:
\begin{eqnarray}
\label{degenerated2F1_z1}
&&\Fh21\Fe{a,b}{a+b+m} \frac{1}{\Gamma(a+b+m)}
=
\nonumber \\
&&\frac{\Gamma(m)}{\Gamma(a+m)\Gamma(b+m)}
\sum_{n=0}^{m-1} \frac{(a)_n(b)_n}{(1-m)_n n!}(1-z)^n
\nonumber \\
&&+\frac{(1-z)^m(-1)^m}{\Gamma(a)\Gamma(b)}
\sum_{n=0}^{\infty}\frac{(a+m)_n (b+m)_n}{(n+m)!n!}
[h_n^{''}-\ln(1-z)](1-z)^n,
\\
&& \nonumber \\
&& -\pi < \arg (1-z)<\pi,\qquad a,b \neq 0,-1,2,{\ldots} ,
\nonumber
\end{eqnarray}
where
\begin{equation}
h_n^{''}=\psi(n+1)+\psi(n+m+1)-\psi(a+n+m)-\psi(b+n+m),
\end{equation}
and
\begin{eqnarray}
\label{degenerated2F1_zinf}
&&\Gamma(a+m)[\Gamma(c)]^{-1}\left._2F_1\right.(a,a+m;c;z)
\nonumber \\
&&~~~~~~~~~=\frac{(-z)^{-a-m}}{\Gamma(c-a)}
\sum_{n=0}^{\infty}\frac{(a)_{m+n}(1-c+a)_{n+m}}
{n!(n+m)!} z^{-n}[\ln(-z)+h_n]
\nonumber \\
&&~~~~~~~~~~~~~~~
+(-z)^{-a}\sum_{n=0}^{m-1}\frac{(a)_n\Gamma(m-n)}
{\Gamma(c-a-n)n!}z^{-n},
\\
&&\nonumber \\
&& |\arg(-z)|<\pi,\qquad
a \neq 0,-1, -2, \ldots,\qquad
m=0,1, \ldots,
\nonumber
\end{eqnarray}
where
\begin{equation}
h_n=\psi(1+m+n)+\psi(1+n)-\psi(a+m+n)-\psi(c-a-m-n).
\end{equation}
Here it is understood that the sum $\sum_0^{m-1}$ is empty for $m=0$.

\boldmath
\subsection{Useful formulae for the Appell function $F_3$}
\unboldmath

The Appell function $F_3$ has the following series representations:
\begin{eqnarray}
F_3(\alpha,\alpha',\beta,\beta',\gamma;x,y)&=&
\sum_{m=0}^{\infty}\sum_{n=0}^{\infty}
\frac{(\alpha)_m (\alpha')_n (\beta)_m (\beta')_n}
{(\gamma)_{m+n}} \frac{x^m}{m!}\frac{y^n}{n!}
\nonumber \\
&=&
\sum_{n=0}^{\infty}\frac{(\alpha')_n (\beta')_n}
{(\gamma)_n~n!} y^n~
\Fh21\Fx{\alpha,\beta}{\gamma+n}
\nonumber \\
&=&
\sum_{n=0}^{\infty}\frac{(\alpha)_n (\beta)_n}
{(\gamma)_n~n!} x^n~
\Fh21\Fy{\alpha',\beta'}{\gamma+n}.
\end{eqnarray}
The integral representation of the $F_3$ function used for the
derivation of the one-fold integral representations of the $\Phi$ function
reads:
\begin{equation}
F_3(\alpha,\alpha',\beta,\beta',\gamma; x, y)=
\frac{\Gamma(\gamma)}{\Gamma(\gamma-\beta)
\Gamma(\beta)}
\int_0^1\frac{u^{\gamma-\beta-1}(1-u)^{\beta-1}}
{(1-x+ux)^{\alpha}}
\Fh21\Fuy{\alpha',\beta'}{\gamma-\beta}~du.
\label{intrepF3inF21}
\end{equation}
This integral representation follows from Eq.~(20) in
Ref.~\cite{Prudnikov:1986}.

\section{Appendix B}

In this appendix, we present the system of differential equations for the
integral $I_5^{(d)}$ to be used for the derivation of the differential
equations for the periodic constants $P_p(d)$ and $P_m(d)$. 
To obtain this system of differential equations for $I_5^{(d)}$,  we exploit
the method described in Ref.~\cite{Tarasov:1996bz}.
According to this method, derivatives with respect to $s_{ij}$ may be expressed
in terms of integrals with shifted space-time dimensions.
Explicit expressions for such derivatives may be derived from the integral
representation with $\alpha$ parameters,
\begin{equation}
I_5^{(d)}(s_{13},s_{14},s_{24},s_{25},s_{35})
= \frac{1}{i^{(d+2)/2}} \int\limits_0^{\infty} d\alpha_1
{\ldots} \int\limits_0^{\infty} d\alpha_5
\frac{\exp\{i \frac{Q}{D} \}}{D^{\frac{d}{2}}},
\label{eq:alpha}
\end{equation}
where
\begin{eqnarray}
Q&=&\alpha_1\alpha_3 s_{13}
+\alpha_1\alpha_4s_{14}
+\alpha_2\alpha_4 s_{24}
+\alpha_2 \alpha_5s_{25}
+\alpha_3\alpha_5s_{35},
\nonumber \\
D &=& \alpha_1+\alpha_2+\alpha_3+\alpha_4+\alpha_5.
\end{eqnarray}
From Eq.~(\ref{eq:alpha}), it follows that
\begin{eqnarray}
&&
\frac{\partial I_5^{(d)}(s_{13},s_{14},s_{24},s_{25},s_{35})}{\partial s_{13}}
=\int \frac{d^{d+2}q}{i \pi^{\frac{d+2}{2}}} \frac{P}{D_1 D_3},
\nonumber \\
&&
\frac{\partial I_5^{(d)}(s_{13},s_{14},s_{24},s_{25},s_{35})}{\partial s_{14}}
=
\int \frac{d^{d+2}q}{i \pi^{\frac{d+2}{2}}} \frac{P}{D_1D_4},
\nonumber \\
&&
\frac{\partial I_5^{(d)}(s_{13},s_{14},s_{24},s_{25},s_{35})}
{\partial s_{24}}
=\int \frac{d^{d+2}q}{i \pi^{\frac{d+2}{2}}} \frac{P}{D_2D_4},
\nonumber \\
&&
\frac{\partial I_5^{(d)}(s_{13},s_{14},s_{24},s_{25},s_{35})}{\partial s_{25}}
=\int \frac{d^{d+2}q}{i \pi^{\frac{d+2}{2}}} \frac{P}{D_2D_5},
\nonumber \\
&&
\frac{\partial I_5^{(d)}(s_{13},s_{14},s_{24},s_{25},s_{35})}{\partial s_{35}}
=\int \frac{d^{d+2}q}{i \pi^{\frac{d+2}{2}}} \frac{P}{D_3D_5},
\label{derivs_via_dp2ints}
\end{eqnarray}
where
\begin{equation}
P=\frac{1}{D_1D_2D_3D_4D_5},
\end{equation}
and $D_j$ are defined in Eq.~(\ref{propagators}).
In order to reduce the $(d+2)$-dimensional integrals on the right-hand sides of
the relations in Eq.~(\ref{derivs_via_dp2ints}) to a set of basic integrals, we
use recurrence relations \cite{Fleischer:1999hq}.
All calculations are performed with the help of computer program package Maple.
The resulting 5 differential equations for the integral $I_5^{(d)}$ all have
the form of Eq.~(\ref{dif_equ_for_i5}).
The polynomials $R_{ij}^{(k)}$ and $\phi_{ij}$ occuring therein are found to be
\begin{eqnarray}
R^{(1)}_{13}&=&
(-s_{25}+s_{24}+s_{35})(s_{14} s_{25}-s_{35} s_{14}-s_{13} s_{25}+s_{24} s_{35}+s_{13} s_{24}),
\nonumber \\
R^{(2)}_{13}&=&
-(s_{14}^2 s_{25}-s_{14}^2 s_{35}+s_{35} s_{14} s_{25}-s_{13} s_{14} s_{25}
+2 s_{13} s_{35} s_{14}-s_{14} s_{35}^2+s_{13} s_{24} s_{14}
\nonumber \\
&&{}+s_{24} s_{35} s_{14}
+s_{25} s_{13} s_{35}+s_{24} s_{35}^2-s_{24} s_{13} s_{35}),
\nonumber \\
R^{(3)}_{13}&=&
-(-s_{24} s_{35}+s_{35} s_{14}+s_{13} s_{24}-s_{13} s_{25}-s_{14} s_{25})(s_{14}+s_{25}-s_{24}),
\nonumber \\
R^{(4)}_{13}&=&
s_{13} (-s_{14} s_{25}+s_{35} s_{14}-2 s_{25}^2+2 s_{24} s_{25}-s_{13} s_{25}+2 s_{25} s_{35}+
s_{13} s_{24}-s_{24} s_{35}),
\nonumber \\
R^{(5)}_{13}&=&
-s_{13} (s_{14} s_{25}-s_{35} s_{14}-2 s_{14} s_{24}-s_{13} s_{25}-2 s_{24} s_{25}+s_{13} s_{24}+s_{24} s_{35}+2 s_{24}^2),
\nonumber \\
R^{(1)}_{14}&=&
(s_{13} s_{25}-s_{13} s_{24}+s_{24} s_{35}+s_{35} s_{14}-s_{14} s_{25})(s_{24}+s_{35}-s_{25}),
\nonumber \\
R^{(2)}_{14}&=&
s_{14} (s_{14} s_{25}-s_{24} s_{35}+2 s_{13} s_{35}-s_{35} s_{14}-s_{13} s_{25}
+s_{13} s_{24}+2 s_{25} s_{35}-2 s_{35}^2),
\nonumber \\
R^{(3)}_{14}&=&
s_{14} ( s_{35} s_{14}-s_{14} s_{25}-2 s_{25}^2+2 s_{24} s_{25}-s_{13} s_{25}+2 s_{25} s_{35}
+s_{13} s_{24}-s_{24} s_{35}),
\nonumber \\
R^{(4)}_{14}&=&
(s_{35}-s_{13}-s_{25})(s_{35} s_{14}-s_{24} s_{35}+s_{13} s_{24}-s_{13} s_{25}-s_{14} s_{25}),
\nonumber \\
R^{(5)}_{14}&=&
(s_{13} s_{24}^2-s_{24} s_{13} s_{35}+s_{24} s_{35} s_{14}-2 s_{13} s_{24} s_{14}
-s_{24} s_{13} s_{25}-s_{35} s_{24}^2
\nonumber \\
&&{}+s_{13}^2 
s_{24}-s_{24} s_{14} s_{25}+s_{13} s_{14} s_{25}-s_{13} s_{35} s_{14}-s_{13}^2 s_{25}),
\nonumber \\
R_{24}^{(1)}&=&
s_{24} (s_{14} s_{25}-s_{24} s_{35}+2 s_{13} s_{35}-s_{35} s_{14}
-s_{13} s_{25}+s_{13} s_{24}+2 s_{25} s_{35}-2 s_{35}^2),
\nonumber \\
R_{24}^{(2)}&=&
(-s_{13} s_{25}+s_{13} s_{24}-s_{24} s_{35}-s_{35} s_{14}
+s_{14} s_{25})(-s_{35}-s_{14}+s_{13}),
\nonumber \\
R_{24}^{(3)}&=&
-(-s_{14}^2 s_{25}+s_{14}^2 s_{35}-s_{14} s_{25}^2+s_{13} s_{14} s_{25}+s_{35} s_{14} s_{25}
\nonumber \\
&&{}+2 s_{24}
 s_{14} s_{25}+s_{13} s_{24} s_{14}-s_{24} s_{35} s_{14}+s_{13} s_{25}^2+s_{24} s_{35} s_{25}-s_{24} s_{13} s_{25}),
\nonumber \\
R_{24}^{(4)}&=&
(s_{13}+s_{25}-s_{35}) (s_{35} s_{14}+s_{13} s_{24}-s_{24} s_{35}
-s_{14} s_{25}+s_{13} s_{25}),
\nonumber \\
R_{24}^{(5)}&=&
-s_{24} (-s_{24} s_{35}+s_{13} s_{24}-2 s_{13} s_{14}-2 s_{13} s_{35}
+s_{13} s_{25}+2 s_{13}^2-s_{14} s_{25}+s_{35} s_{14}),
\nonumber \\
R_{25}^{(1)}&=&
-(-s_{24} s_{35}^2+2 s_{24} s_{35} s_{25}-s_{24} s_{13} s_{25}+s_{25} s_{13} s_{35}
\nonumber \\
&&{}+s_{24} s_{13} s_{35}+s_{13} s_{24}^2+s_{24} s_{14} s_{25}
-s_{35} s_{24}^2+s_{14} s_{35}^2-s_{35} s_{14} s_{25}+s_{24} s_{35} s_{14}),
\nonumber \\
R_{25}^{(2)}&=&
(-s_{13} s_{25}+s_{13} s_{24}+s_{35} s_{14}-s_{24} s_{35}+s_{14} s_{25})(s_{35}+s_{14}-s_{13}),
\nonumber \\
R_{25}^{(3)}&=&
s_{25} (-2 s_{14}^2-s_{14} s_{25}+2 s_{13} s_{14}-s_{35} s_{14}
+2 s_{14} s_{24}+s_{13} s_{25}-s_{13} s_{24}+s_{24} s_{35}),
\nonumber \\
R_{25}^{(4)}&=&
-s_{25} (-s_{24} s_{35}+s_{13} s_{24}-2 s_{13} s_{14}-2 s_{13} s_{35}+s_{13} s_{25}+2 s_{13}^2-s_{14} s_{25}
+s_{35} s_{14}),
\nonumber \\
R_{25}^{(5)}&=&
(s_{13}+s_{24}-s_{14})(s_{35} s_{14}+s_{13} s_{24}-s_{24} s_{35}-s_{14} s_{25}+s_{13} s_{25}),
\nonumber \\
R_{35}^{(1)}&=&
-s_{35} (s_{14} s_{25}-s_{35} s_{14}-2 s_{14} s_{24}-s_{13} s_{25}-2 s_{24} s_{25}
+s_{13} s_{24}+s_{24} s_{35}+2 s_{24}^2),
\nonumber \\
R_{35}^{(2)}&=&
s_{35}(-2 s_{14}^2-s_{14} s_{25}+2 s_{13} s_{14}
-s_{35} s_{14}+2 s_{14} s_{24}+s_{13} s_{25}-s_{13} s_{24}
+s_{24} s_{35}),
\nonumber \\
R_{35}^{(3)}&=&
(s_{14}+s_{25}-s_{24}) (-s_{13} s_{25}+s_{13} s_{24}
+s_{35} s_{14}-s_{24} s_{35}+s_{14} s_{25}),
\nonumber \\
R_{35}^{(4)}&=&
-(s_{14} s_{25}^2-s_{35} s_{14} s_{25}+s_{13} s_{14} s_{25}+s_{13} s_{35} s_{14}
-s_{13} s_{25}^2
-s_{13}^2 s_{25}
\nonumber \\
&&{}+s_{24} s_{35} s_{25}+s_{24} s_{13} s_{25}+2 s_{25} s_{13} s_{35}
-s_{24} s_{13} s_{35}+s_{13}^2 s_{24}),
\nonumber \\
R_{35}^{(5)}&=&
(s_{14} s_{25}-s_{35} s_{14}-s_{13} s_{25}+s_{24} s_{35}+s_{13} s_{24})(s_{13}+s_{24}-s_{14}),
\label{R_polynoms}
\end{eqnarray}
and
\begin{eqnarray}
\phi_{13} &=& -3 s_{14}^2 s_{25}^2+6 s_{25} s_{14} (s_{14}-s_{24}) s_{35}
-s_{14} s_{25} (s_{24}-s_{25}) s_{13}
\nonumber \\
&&{}-3 (s_{14}-s_{24})^2 s_{35}^2
+2 (s_{24}-s_{25})^2 s_{13}^2
+(s_{24}^2-s_{14} s_{25}-s_{14} s_{24}-s_{24} s_{25}) s_{13} s_{35},
\nonumber \\
\phi_{14} &=& -3 s_{13}^2 s_{25}^2+(s_{35}^2-s_{25} s_{35}
-s_{13} s_{35}-s_{13} s_{25}) s_{14} s_{24}
\nonumber \\
&&{}+s_{13} s_{25} (s_{25}-s_{35}) s_{14}+6 s_{13} s_{25} (s_{13}-s_{35}) s_{24}
-3 (s_{13}-s_{35})^2 s_{24}^2+2 (s_{25}-s_{35})^2 s_{14}^2,
\nonumber \\
\phi_{24} &=& -3 s_{14}^2 s_{35}^2-6 s_{35} s_{14} (s_{13}-s_{14}) s_{25}
-s_{35} s_{14} (s_{13}-s_{35}) s_{24}
\nonumber \\
&&{}+(s_{13}^2-s_{35} s_{14}-s_{13} s_{14}-s_{13} s_{35}) s_{24} s_{25}
+2 (s_{13}-s_{35})^2 s_{24}^2-3 (s_{13}-s_{14})^2 s_{25}^2,
\nonumber \\
\phi_{25} &=& -3 s_{14}^2 s_{35}^2-s_{35} s_{14} (s_{13}-s_{14}) s_{25}
-6 s_{35} s_{14} (s_{13}-s_{35}) s_{24}
\nonumber \\
&&{}+(s_{13}^2-s_{35} s_{14}-s_{13} s_{14}-s_{13} s_{35}) s_{24} s_{25}
-3 (s_{13}-s_{35})^2 s_{24}^2+2 (s_{13}-s_{14})^2 s_{25}^2,
\nonumber \\
\phi_{35} &=& -3 s_{14}^2 s_{25}^2+s_{25} s_{14} (s_{14}-s_{24}) s_{35}
-6 s_{14} s_{25} (s_{24}-s_{25}) s_{13}+2 (s_{14}-s_{24})^2 s_{35}^2
\nonumber \\
&&{}-3 (s_{24}-s_{25})^2 s_{13}^2
+(s_{24}^2-s_{14} s_{25}-s_{14} s_{24}-s_{24} s_{25}) s_{13} s_{35},
\label{phi_polynoms}
\end{eqnarray}
respectively.

\end{document}